\documentclass[manuscript]{aastex}
\usepackage{outlines}
\usepackage{enumitem}
\usepackage{float}
\usepackage{natbib}

\slugcomment{to appear in a AAS Journal}

\shorttitle{Model Light curves}
\shortauthors{Basri \& Shah}

\begin{document}


\title{The Information Content in Analytic Spot Models \break
of Broadband Precision Light Curves. II. \break 
 Spot Distributions and Lifetimes,  \break
 Global and Differential Rotation}


\author{Gibor Basri and Riya Shah}

\affil{Astronomy Department, University of California,
    Berkeley, CA 94720}

\email{basri@berkeley.edu} 





\begin{abstract}

With the advent of space-based precision photometry missions the quantity and quality of starspot light curves has greatly increased. This paper presents a large number of starspot models and their resulting light curves to: 1) better determine light curve metrics and methods that convey useful physical information, 2) understand how the underlying degeneracies of the translation from physical starspot distributions to the resulting light curves obscure that information. We explore models of relatively active stars at several inclinations while varying the number of (dark) spots, random spot distributions in position and time, timescales of growth and decay, and differential rotation. We examine the behavior of absolute and differential variations of individual intensity dips and overall light curves, and demonstrate how complex spot distributions and behaviors result in light curves that typically exhibit only one or two dips per rotation. Unfortunately simplistic ``one or two spot" or ``active longitude" descriptions or modeling of light curves can often be highly misleading. We also show that short ``activity cycles" can easily be simply due to random processes.

It turns out to be quite difficult to disentangle the competing effects of spot lifetime and differential rotation, but under most circumstances spot lifetime is the more influential of the two. Many of the techniques tried to date only work when spots live for many rotations. These include autocorrelation degradation for spot lifetimes and periodograms for both global and differential rotation. Differential rotation may be nearly impossible to accurately infer from light curves alone unless spots live for many rotations. The Sun and solar-type stars its age or older are unfortunately the most difficult type of case. Further work is needed to have increased confidence in light curve inferences.

\end{abstract}





\section{Introduction\label{sec:intro}}

Starspots are regions of a star's photosphere in which concentrated magnetic fields suppress local convection and prevent hotter material from rising as efficiently. This makes starspots cooler and therefore darker than surrounding areas. Their presence manifests as a deficit in a star's brightness and thus a dip in its light curve, easily recognisable due to characteristic timescales set by the stellar rotation. Although spots have been observed and studied over centuries on our Sun, insights into the behavior of sunspots are not necessarily transferable to other stars. It is not clear to what extent the solar analogy works for more active stars, since they often have larger spotted areas whose lifetimes may be longer. Sunspots preferentially appear as bipolar pairs at particular latitudes (which drift toward the equator over the solar cycle), while on more active stars larger spotted regions can appear at a larger range of latitudes. There is a tendency for spots to be found preferentially in polar regions on rapidly rotating stars \citep{Stras09}.    

Understanding the properties of starspots on various types of stars could lead to important insights into magnetic dynamos, organization and evolution of surface fields, and stellar surface differential rotation. Missions such as CoRoT \citep{Auv09} and Kepler \citep{Bor10} have provided a wealth of high-precision photometric data over long time periods; a boon for stellar researchers. Substantial work has been done on modeling starspots using light curves gathered by space-based telescopes. For example, \citet{Moss09} and \citet{Lanz09} created analytic spot models based on data collected by CoRoT to determine stellar properties such as rigid-body rotation, spot lifetimes, and differential rotation. The primary Kepler mission gathered data on over one hundred thousand main sequence stars nearly continuously for four years. The data provided by these missions and others have led to much research on the morphological properties of stellar light curves, including \citet{Deg12,RRB13,Niel15,Sant17,Ark18} among many other papers. Many of the Kepler stars with measured rotation periods are more active and have more photometric variability than our Sun \citep{Bas10} because that makes their periods easier to measure. It is that class of star this paper is most relevant to.  

In this paper we analyze a plethora of starspot models based on the methodologies described in \citet{Walk13}; see that previous paper in this series for a fuller introduction. We now conduct a more physical analysis; the new models contain different combinations of astrophysically interesting parameters including spot numbers, frequency of appearance, differential rotation, spot lifetimes, and stellar inclination. Such models provide thousands of noiseless data samples which generate aggregate statistics that can help us to understand how much information is extricable from precision light curves with excellent time coverage. We focus on morphological characteristics of light curves such as the fraction of time spent in single or double intensity dip modes (per rotation), variability in the depth and duration of individual dips, the overall range of light curve amplitudes, variations in overall light deficit per rotation (spot coverage), timescales of variations, and periodogram and autocorrelation statistics. We concentrate on what might be considered the ``worst case": fully random distributions of spots in location and time. It is possible that real stars exhibit more systematic spot distributions that would sometimes mitigate some of the issues we uncover.

Papers to date that focus on starspot properties derived from stellar light curves contain severe simplifications about starspot numbers, distributions, and evolution. For example \citet{Lanz14} and \citet{Name19} associate double dips (per rotation) in light curves with the appearance of two distinct (large) starspots. Although this assumption has a long history \citep{Rod86}, it is questionable. Double dips can be caused by any number of starspots greater than one, single dips can be caused by one or more. It is usually not possible to determine how many spots actually are present on a star's surface (or where they are) from just a light curve, a point central to this paper. We illustrate how light curves often greatly simplify underlying complexity, and reinforce that solutions to the inverse problem are highly degenerate. This has been known for a long time, but the question is to what extent that masks or distorts the information one seeks to learn about starspots and stellar magnetic fields. This paper examines patterns within certain light curve metrics to see to how much unique information they really provide on the spot distributions, lifetimes, differential rotation, and periodicity of active stars. 

The paper is organized as follows. Section \ref{sec:Modeling} describes our spot modeling method, parameter sets for these models, and the metrics we employ to characterize light curve behaviors. Section \ref{sec:Results} contains the main results from our extensive model testing on metrics including the single/double ratio (SDR), a full rotation coverage metric, and variability statistics for individual dips and full light curves for models with varying numbers of spots, spot lifetimes, and stellar inclinations, with and without differential rotation. In Section \ref{sec:Conclusions} we discuss how these results impact research that has been done to infer starspot properties from light curves, explain the impact of degeneracies in light curves, and  re-examine a variety of methods and conclusions that are in the literature to date. Our purpose is not to criticize what has been done, but rather to advocate why future work should lift some of the simplifications that have previously been made, or better recognize their effects.

\section{Analytic Spot Models \label{sec:Modeling}}

Starspot forward modeling is quite straightforward. One must specify the contrast of features compared to the quiet photosphere, and the location and size (perhaps shape) of them. Then the intensity of the visible hemisphere of the star with such features at various locations is essentially just a problem in geometry. Such computations are quite fast, making it possible to examine many models. One approach is to divide the surface of the star into pixels, and specify the brightness of each pixel. Another is to utilize round spots of various sizes at known locations; the deficit due to each at any given moment can then be analytically computed. That is the approach we use. The basic modeling code is taken from \citet{Walk13}; the underlying scheme of analytic spot modeling is due to \citet{Dor87}. Our IDL code uses a uniform (pseudo-)random number generator to produce distributions of spot positions and birthdates in a given run. The boundary conditions on those two variables are parameters of the model. All the light curves are then generated, stored, and analyzed. Each model consists of 1000-3000 individual trials with the parameters of interest fixed, and we produce a set of models in which those parameters are given a systematic set of different values. We preserve the parameters for each model and allow them to be re-used with different values of stellar inclination or differential rotation. This permits a more direct comparison of what happens when those two variables are altered. It would also be possible to compute the same model with various implementations of spot evolution functions.

\subsection{Model Parameters \label{sec:Parameters}}

The modeling procedure has many relevant parameters that must be set, but the most important are the number of spots, their spatial distribution, how much (if any) differential rotation is present, what lifetime the spots have (or no evolution), and the inclination at which the star is observed. We use combinations of these parameters to test many different cases. Below we describe the different parameters employed. Almost all the trials in this paper contain 3000 runs for a given parameter set. The number of spots present on the star at a given time is fixed throughout a run if there is no spot evolution but varies somewhat if there is evolution, because spot birthdates are randomized. Spot positions are randomized within the constraints of the spatial parameters for each separate run of the same parameter set. The timescale for the models is specified in rotation periods (this can be scaled to any actual rotation period by multiplying by the period in real time units). We ran the models for 100 rotations using a time resolution of 30 evenly spaced points per rotation.

One essential parameter is the location (stellar longitude and latitude) of each starspot. Our modeling program allows us to dictate a range of longitudes and latitudes (in degrees) within which the locations of starspots are randomized. The defaults for these parameters encompass all visible latitudes on the star (spots are not placed at latitudes that will never be visible for the initial inclination) and all longitudes. We use uniform random distributions to accomplish this. If it is desired that the whole star be available for spots the initial inclination is set to 90$^\circ $; our procedure allows one to subsequently view the same generated case from different inclinations. Options to restrict the latitudes of the spots to latitude belts of specified width (symmetric about the equator) to mimic the behavior of sunspots are included. We also include an option to restrict spot longitudes to a couple of stripes with set mid-points and widths for the purpose of studying ``active longitudes'', but both of these options are largely reserved for a future paper in order to keep the current analysis simpler. We discuss below how we treat differential rotation, which can change the longitude of a spot over time.

The sizes (radii in degrees) of the starspots must be specified. When a range of sizes is specified, then larger spots dominate the overall spot coverage of the star and the smaller spots are less important for most metrics of the light curve, unless they are much more numerous. After making a few tests we decided to use a single fixed size for the tests in this paper to simplify the analysis. The reason is that with spot evolution, the actual size distribution of spots is constantly changing; the specified spot size is actually just the maximum boundary condition. The results are easily extended to the case when there are lots of small spots and only a few large ones, but that is not so different in practice. In principle, the contrast of a spot could also be varied relative to the quiet star (and one could even include umbral and penumbral parameters). In practice this is a refinement not needed in this study (being degenerate with the spot size distribution); we fix the spot contrast at 0.7 of the photosphere. We have not included the effects of faculae because we do not think they affect Kepler differential light curves \citep{Bas18b}, and they are unlikely to change the basic conclusions of this paper.

An important parameter that is a little less intuitive is the ``spot number", which is a boundary condition on how many spots might be present on the star. Not all of the extant spots will always be in the field of view at a given time, some may be on the hidden side of the star. Spots can be permanent, in which case there is a specified fixed number of them and the same spots are present throughout the run, but in almost all our cases they instead have a specified fixed lifetime. Evolving spots are modeled in a very simple way (similar to \citet{Name19}). Spots appear very small at their birthdate, grow linearly for half their lifetime to their specified maximum size, then decay linearly. We realize that this growth and decay scheme is quite simplified and does not reproduce the solar case, but that doesn't affect the results in this paper (based on a few trials with other growth laws). In any case we don't know much about growth and decay patterns on other stars with varying levels of activity. 

The ``spot number" parameter is related to the average number of spots that are typically present; but is actually a parameter in the routine that generates birthdates. Spots are introduced at random birth dates with an average frequency that produces the desired average spot number. In practice this means that the total number of spots produced during the run is the length of the run divided by the lifetime (both in periods), multiplied by the desired spot number. For example, a run of 100 rotations with the spot number chosen to be 6 and a specified lifetime of 10 rotations will have 60 unique spots produced (at various times) during the run, each of which lasts for 10 rotations. Birth dates are allowed to include nearly 1 rotation before the start of the run as well as most of the last rotation so that the star typically has a reasonable set of spots in various stages of evolution. 

\begin{figure}[H]
\begin{center}
\minipage{0.5\textwidth}%
  \includegraphics[width=\linewidth, height=7cm]{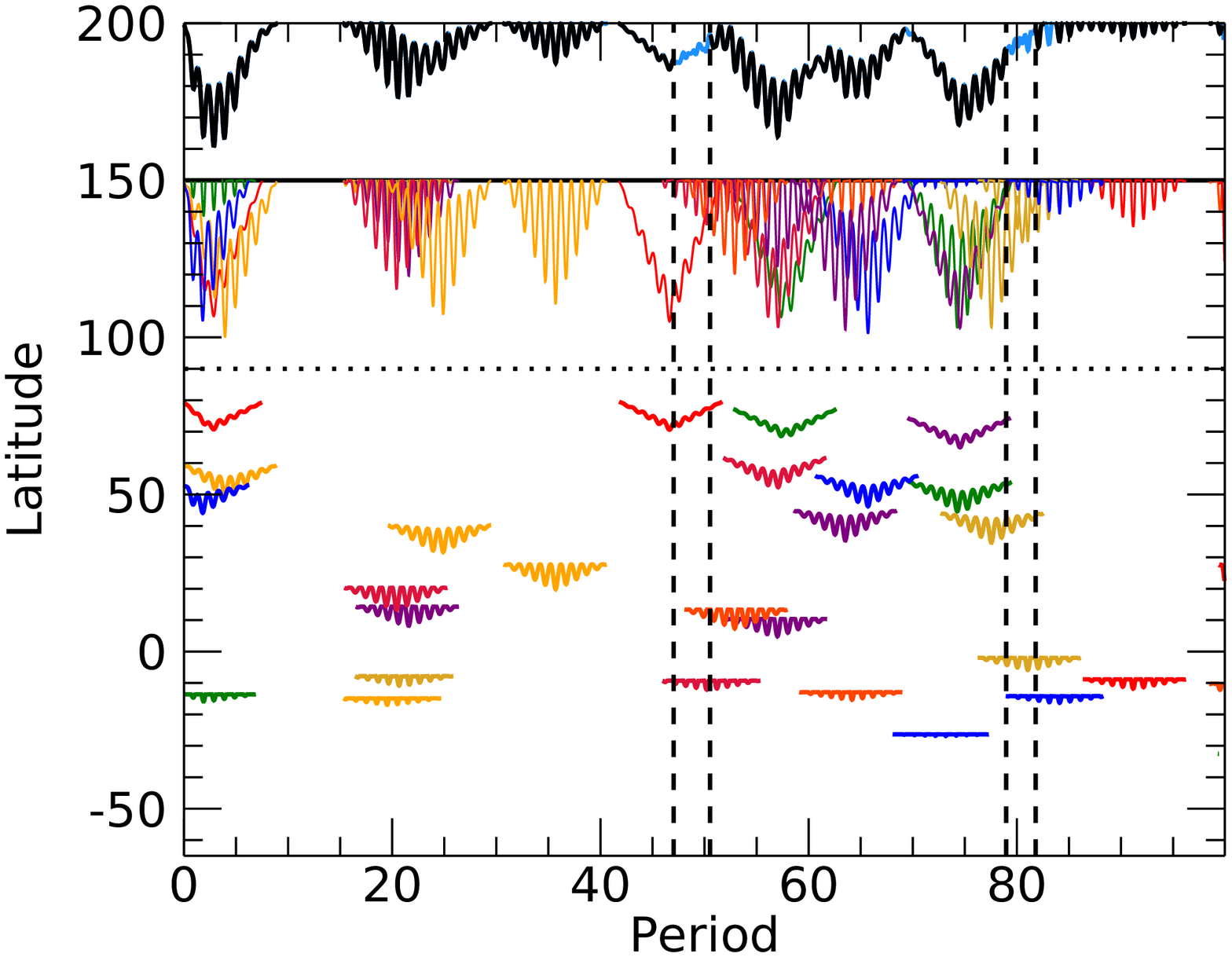}
\endminipage\hfil
\minipage{0.5\textwidth}%
  \includegraphics[width=\linewidth, height=7cm]{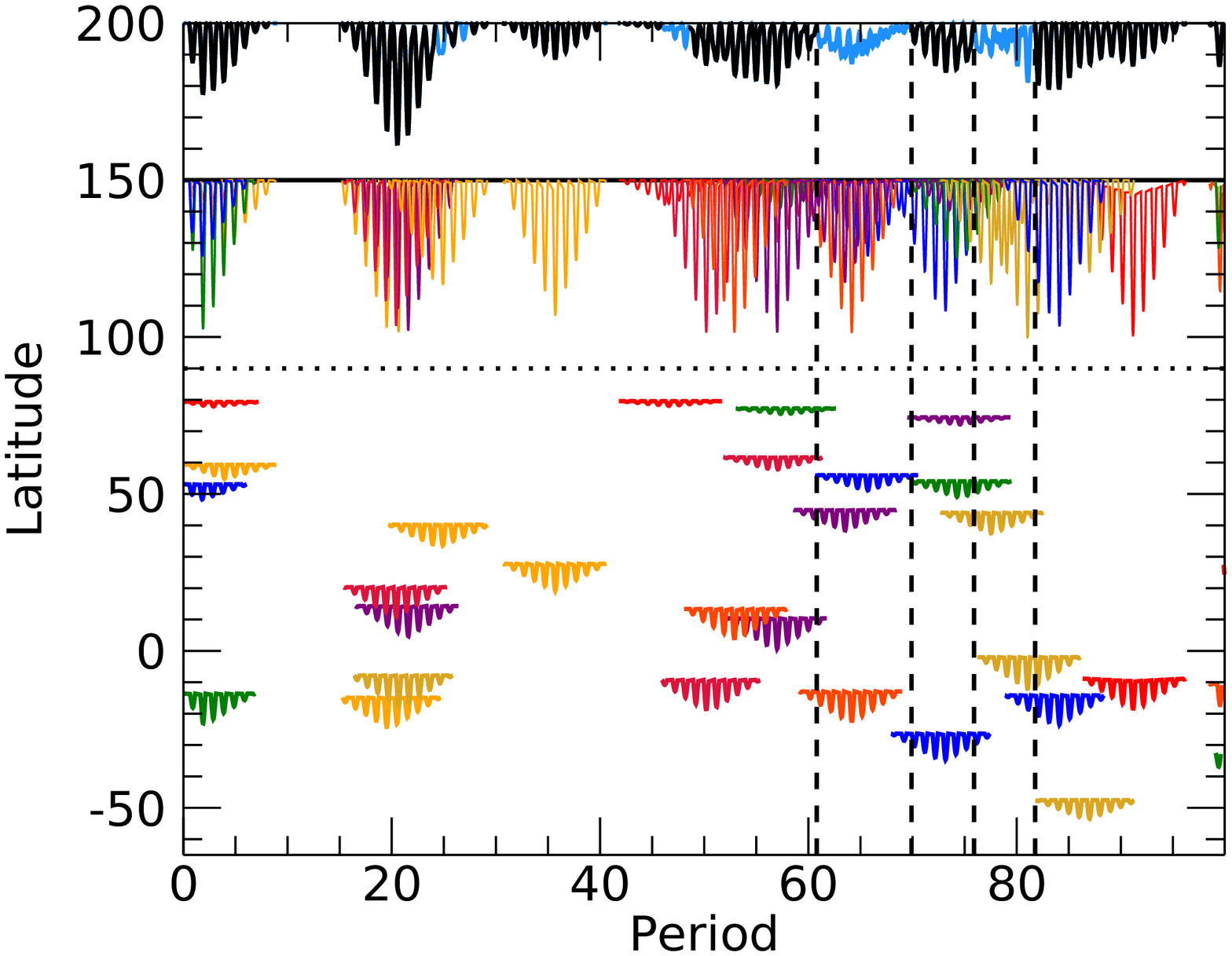}
\endminipage\hfil
\minipage{0.5\textwidth}%
  \includegraphics[width=\linewidth, height=7cm]{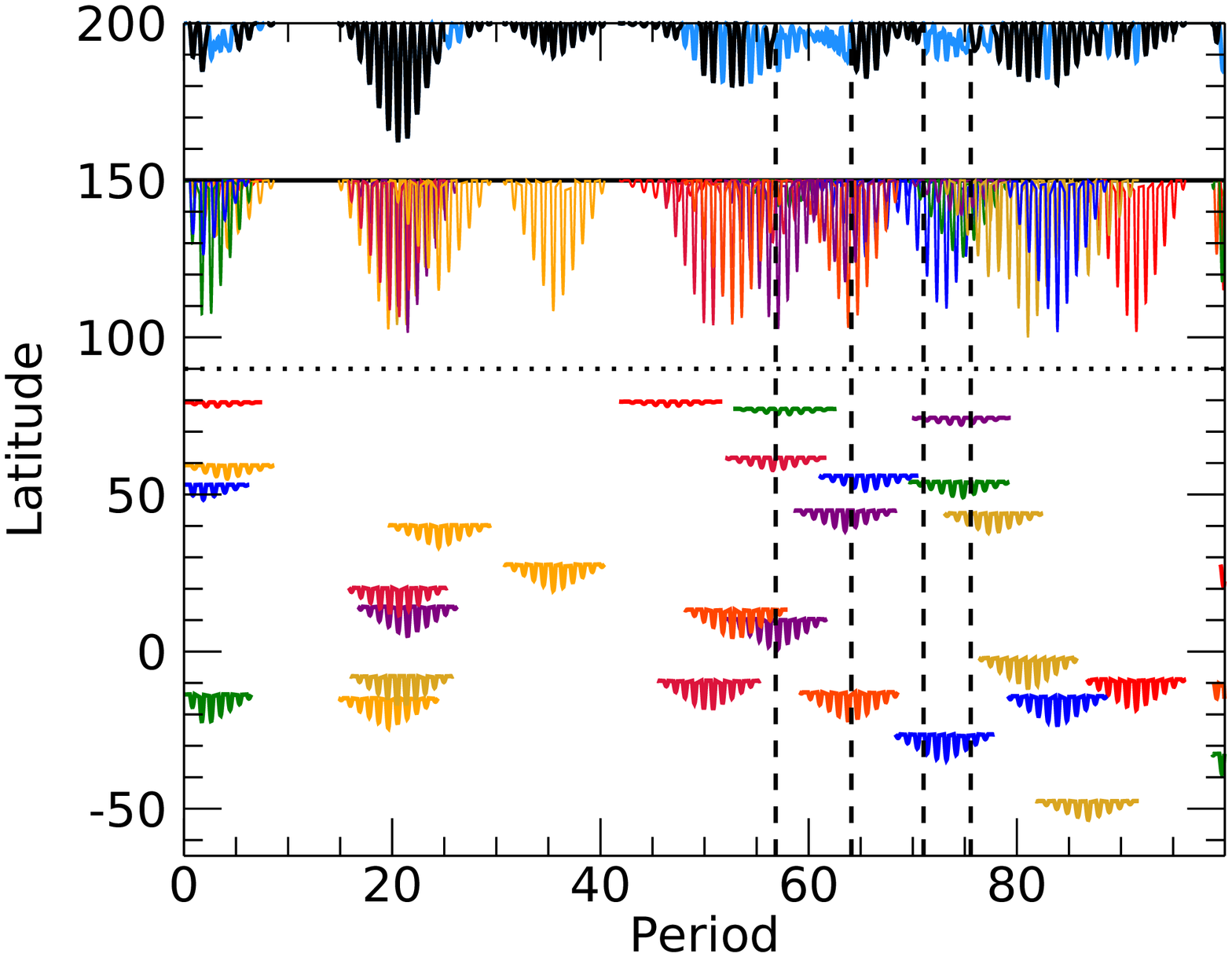}
\endminipage\hfil
\minipage{0.5\textwidth}%
  \includegraphics[width=\linewidth, height=7cm]{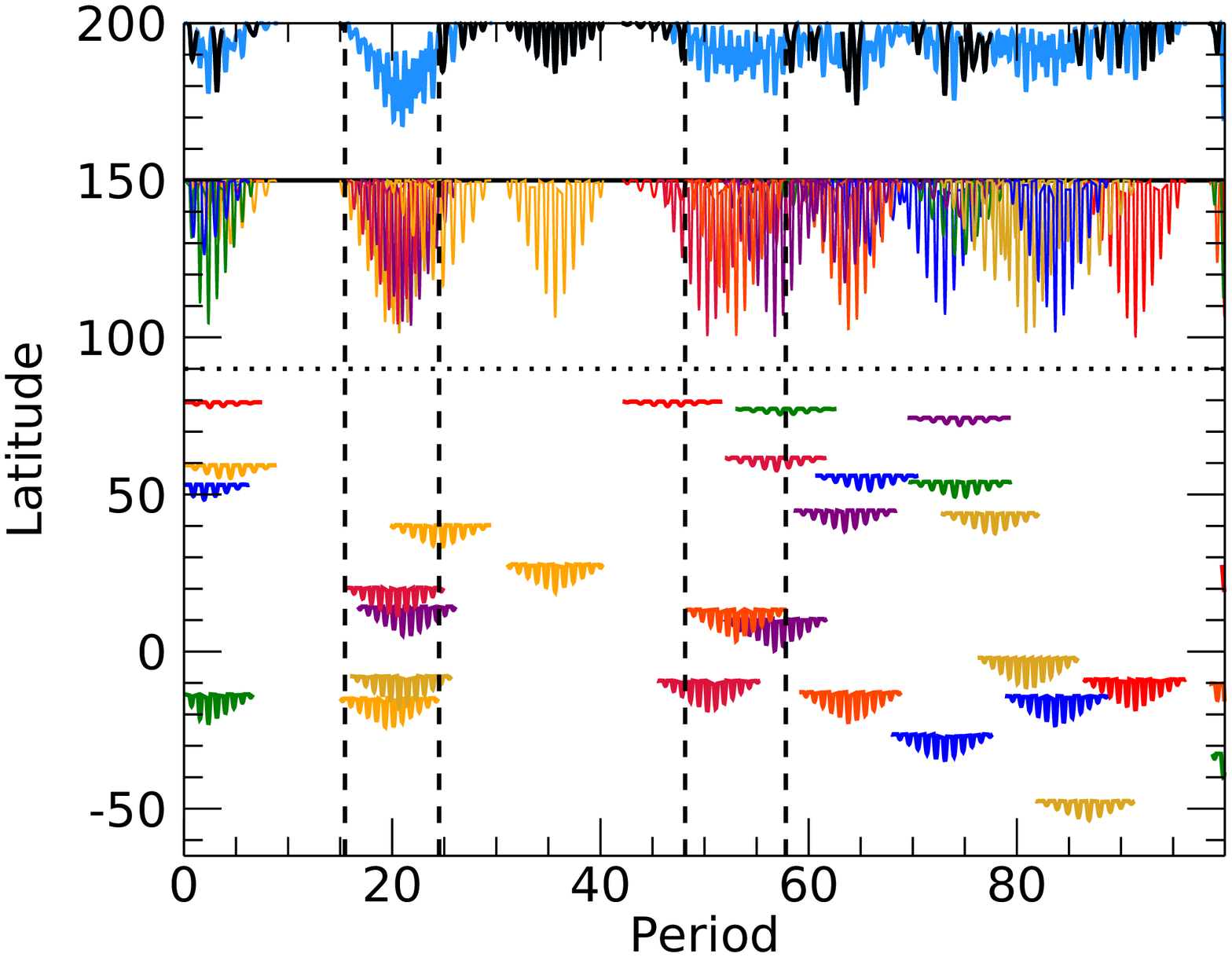}
\endminipage
\caption{Visualizing Spots and Light Curves. All panels have three components from top to bottom: the integrated light curve, overlapping individual light deficit curves, and a visualization of the temporal appearance of the deficits caused by the individual spots at their respective latitudes. The light curves have been scaled to fit so the units are not obvious or given. Black segments on the integrated light curve indicate single-dip segments while blue ones indicate double-dip segments (cf. Section \ref{sec:Metrics}). The top left panel is the original case with spot number 3 and lifetime 10 at $i=30^\circ $. The other three panels are that case viewed from the equator. The top right also has no shear, the bottom left has solar differential rotation, and the bottom right has twice solar shear. In all four panels, the initial spot distributions are the same. The dashed vertical lines bracket the longest two double-dip segments in each panel. }
\label{fig:SameStarDiffIncl}
\end{center}
\end{figure}

We generate an initial set of birthdates distributed randomly within the whole run then introduce an additional randomization of up to two rotations on each birthdate. This process is stochastic, so the actual number of spots present on the star varies somewhat over time around the desired number, and of course not all of the spots present are necessarily visible at a given time as well as having different projected areas in each time step. This resembles the behavior of actual stars, which do not hold the number of spots fixed at all times. For our test cases we used average spot numbers of 3, 6, 9, and 12, with possible spot lifetimes of 1, 2, 5, 10, 20, and 50 rotations. The actual number of spots visible at any given time (with their different and changing projected areas) is of course expressed in the total light deficit at that time. Figure \ref{fig:SameStarDiffIncl} shows an example of how this looks for a particular case with spot number 3. There is a pile-up by chance of 5 spots near rotation 20, following a period of no spots between 10-15. During the pile-up, the star does not recover to the unspotted brightness when viewed with an inclination of $30^\circ $ (upper left), but does at times when viewed at $90^\circ $ (upper right). In both cases the light curve is single-dipped (per rotation) during this time, implying that the spots are somewhat more concentrated on one side of the star. 

It is of great interest to include the possibility of differential rotation in our model because that property is diagnostic of magnetic dynamos and poorly understood for most stars. We adopt the solar latitude law \citep{Snod83} with the option of changing its shear value. For our tests, we used shear values of 0.0 (no differential rotation), 0.2 (solar differential rotation), and 0.4 (twice the solar value). It turns out to be quite useful to implement this by first generating a model with specified spot number, spot birthdates and locations, and spot lifetime. It is then possible to recompute the model with differential rotation by changing the longitudes from their starting values over time as demanded by the shear; one can also view these models at different inclinations. In this way it is possible to directly compare the light curves with and without differential rotation more exactly. 

In Figure \ref{fig:SameStarDiffIncl} for example, the equatorial cases have zero shear in the upper right, solar shear in the lower left, and twice solar shear in the lower right panel. Examining the deep feature near rotation 20 in the no shear case, it is clear that the spots are positioned so that during part of each rotation the light curve recovers to the unspotted value. When solar shear is added, this remains the case, but with twice solar shear the deficit remains substantial throughout those rotations, and the mode changes from one dip per rotation to two. That is because the spots have managed to shift enough in phase to cover all phases with some pattern. 

\begin{figure}[H]
    \begin{center}
    \epsscale{1.0}
    \plotone{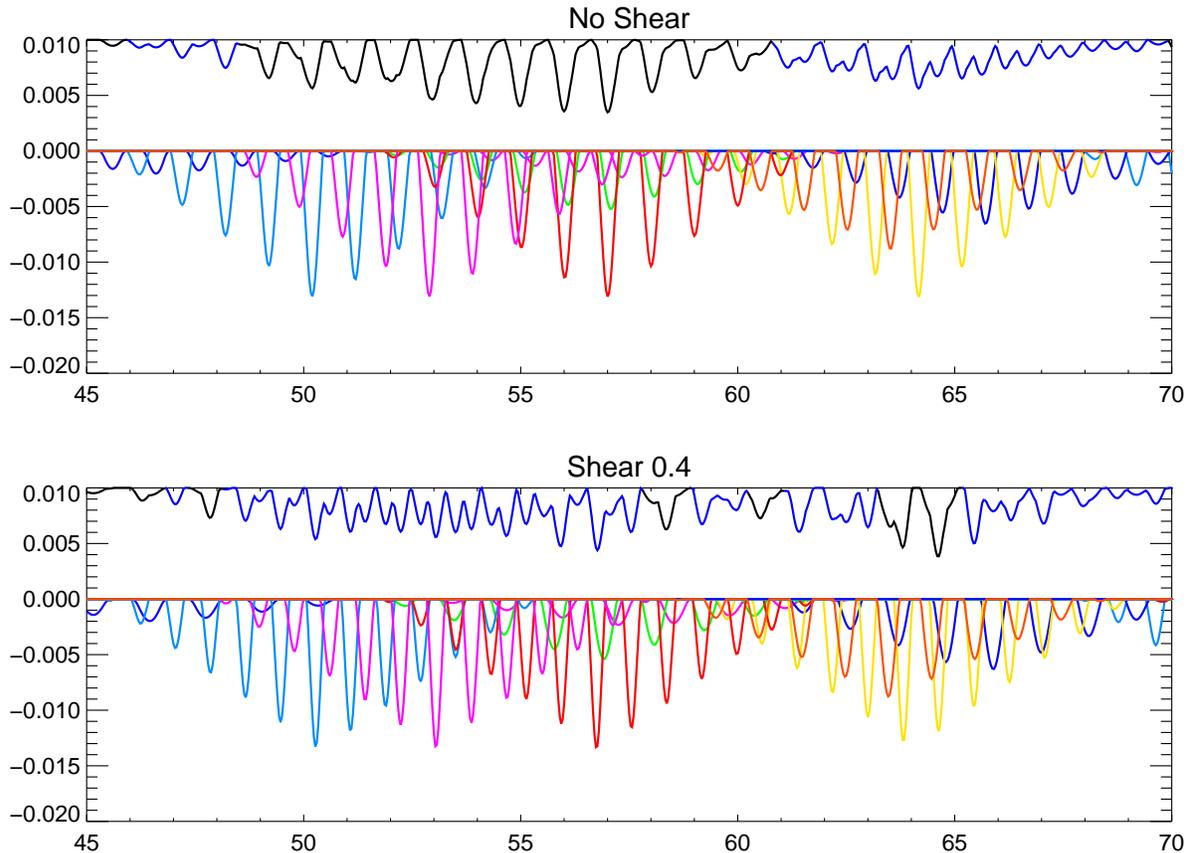}
    \caption{ An expansion of part of the right panels in Figure \ref{fig:SameStarDiffIncl} (spot colors are not the same). The lower curves in each panel are the actual light deficits caused by each spot (which the ordinate provides the scale for), while the upper curves are normalized versions of the total light curve in arbitrary units. The region between rotations 50-60, for example, are converted from single mode to double mode by a shift in the spot phases. }
    \label{fig:OverlapLC}
    \end{center}
\end{figure}

A similar effect can be seen for the segment between 45-70 in the right panels, in which there are transitions from single- to double-dipped as the shear increases (and a bit of the opposite as well). It is easier to see this in a magnified view of the overlapping light curves as given in Figure \ref{fig:OverlapLC}. For example, the segment between 50-55 changes from single to double mode largely because the deeper magenta spot has shifted from having dips adjacent to the blue dips in the upper panel to having them in between the blue dips in the lower panel. A similar interaction occurs between the two magenta spots with help from the green spot from 55-60. One question addressed later in this paper is whether it is possible to distinguish such behavior from similar effects solely due to spot evolution, and whether the actual amount of differential rotation could be measured. 

Finally, an important parameter in how a light curve appears is the stellar inclination. Stars that are pole-on ($i = 0^\circ $) will show no variation (other than due to spot evolution), while spots on stars that are equator-on ($i = 90^\circ $) will be hidden half the time. Inclinations used in our tests were 30$^\circ $, 45$^\circ $,  60$^\circ $, and 90$^\circ $. Figure \ref{fig:SameStarDiffIncl} is an example of how changing the inclination of a star while keeping the spot distributions the same affects the integrated light deficit curve. As can be seen in Figure \ref{fig:SameStarDiffIncl} (upper left panel), when the star is viewed from 30$^\circ $ the deficits for spots at higher latitudes are ``V-shaped" in time; the spots remain constantly visible while growing and decaying. These ``V-shaped" deficits (caused by our linear growth and decay model) produce a larger integrated deficit than for spots which are hidden part of the time (the amount also depends on how close to the sub-observer latitude they are). This tends to increase the total light deficit due to spots in low inclination cases compared with high inclination cases. At $i=90^\circ $ (equator on, other 3 panels), the spots at higher latitudes affect the integrated light deficit curve less while the spots at lower latitudes are more prominent due to area projection effects. In both cases the deficits at lower latitudes are more likely to return to zero for part of each rotation because the spots go behind the star. An example of these effects is seen in the top two panels of Figure \ref{fig:SameStarDiffIncl} between rotations 60-80.

In the case shown in Fig. \ref{fig:SameStarDiffIncl} there are 30 spots in the entire run which live for 10 rotations each, so in one sense there are 3 spots present on average. However, because of the randomization of birthdates it is obvious that the actual number of spots contemporaneously present ranges from 0 to 5. At any given time some of these might be hidden on the back of the star, and their contribution to the light deficit depends on their latitude and phase through the projected area. This is a relatively simple case; on a real star the average number of spots could vary over time and their locations could be constrained in various ways. This example illustrates why it is therefore quite difficult to infer the actual number of spots on a star from its light curve, and the variation of the light curve is due to all these effects. 

We chose a canonical set of model parameters, from which one or more were varied. In this base case the spot location is fully randomized in longitude and visible latitude and the stellar inclination is 60$^\circ $ (close to the statistically  ``most likely" value); this both allows for spots that go behind the visible hemisphere and spots that don't. The maximum spot size is always fixed at 5 degrees in radius. Of course, this size (and spot contrast) could be scaled to produce the depths present in an observed light curve, and in any case the spots grow and decay through smaller sizes. Our canonical spot number is 6. Differential rotation is sometimes subsequently imposed on the model. We do not set a canonical value for the spot lifetime; we always tried a range of lifetimes. In practice, we vary many of the parameters to understand their effects, so the canonical case is just illustrative.

It is important to note that the Sun does not have a light curve that resembles most of the Kepler stars {\it with known rotation periods}. Those stars tend to be as or more variable than the active Sun \citep{Bas13} which makes determining their rotation periods easier but causes an observational bias towards active stars. Recently \citet{Rein20} have shown that when one uses stellar parameters to select stars near to the Sun in characteristics, there is a still a population of stars that are more active than the active Sun; they are mostly the ones whose periods are measured from Kepler light curves. Because the rotation period is essential for any spot analysis, this paper concentrates on spot models that look like those more variable Kepler light curves. The models do not include injection of ``observational" noise so that the underlying effects are clearest. 

\subsection{Light Curve Metrics \label{sec:Metrics}}

In order to characterize a large sample of unique light curves, we utilize several metrics relating to their shapes and temporal behaviors. Light curves of the quality produced by recent precision photometers in space have largely been studied for their periodicity or periodogram appearance \citep{RRB13, McQ14, Niel15, Aig15}, variability amplitude \citep{Bas13, McQ14, Math14, Bas18b}, and more recently their ``single/double" character \citep{Bas18a}. They have also been studied for activity cycles \citep{Rein17, Niel19}, and differential rotation \citep{RRB13,Das16,Sant17}. Some of our metrics are based on the whole light curve, some are defined by the behavior over one rotation period, and some are produced by breaking the light curve into individual ``dips", by which we mean the segments in the light curve between successive local intensity maxima. These dips may last for well under half a rotation period up to most of a rotation period, or sometimes even longer than a rotation period (depending on how the spot distribution and its visibility is evolving).

\begin{figure}[H]
    \begin{center}
    \epsscale{1.0}
    \plotone{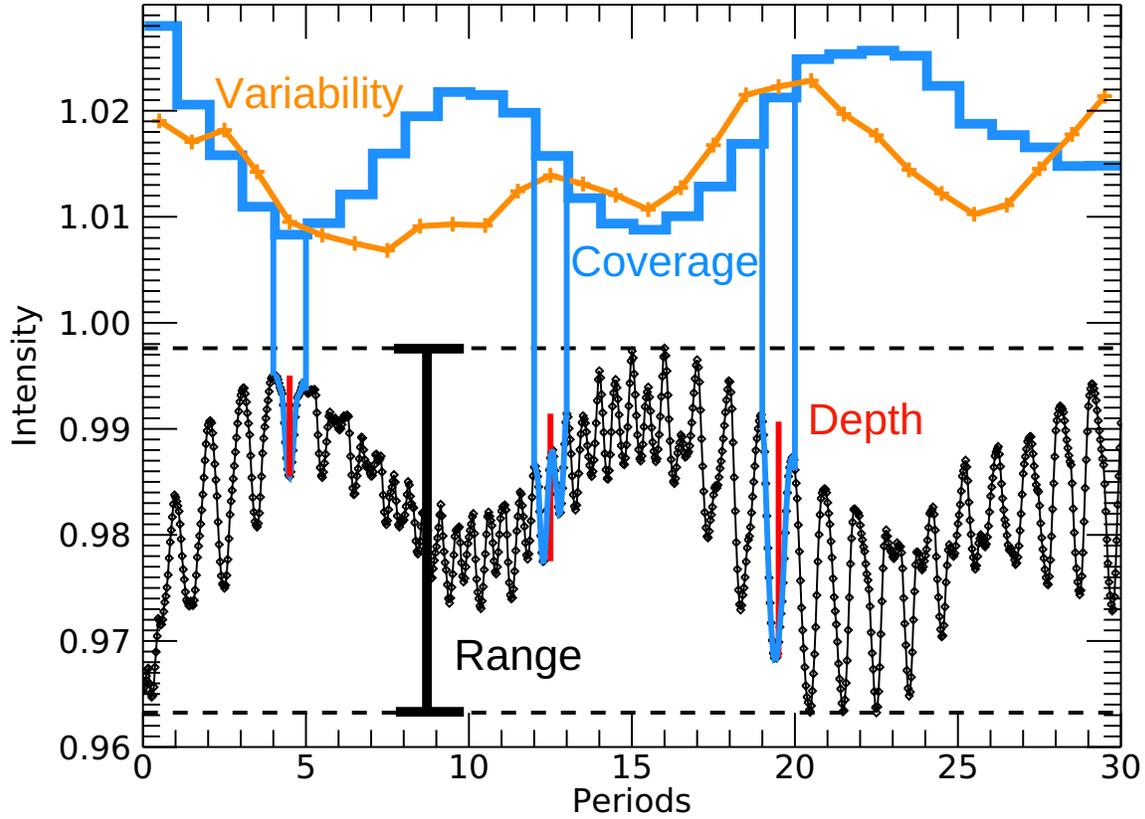}
    \caption{ This figure illustrates how we determine the depth, coverage, variability, and total range metrics.  The bottom half of the figure shows a segment of a light curve from a model star with a spot lifetime of 10 rotations, spot number 12, and $i=60^\circ $. The blue curve shows the spot coverage (integrated deficit over one rotation) while the orange curve shows the variability (time behavior of depth). Specific examples of features that produce points in the depth curve are shown with red vertical lines. Rotation blocks are defined by vertical blue lines. The total range of the light curve is shown by the vertical black line. }
    \label{fig:example_mets}
    \end{center}
\end{figure}

One of the variability metrics we use is the total range, defined as the difference between the highest and lowest intensity data points over the whole light curve. We also define the ``median dip range" as the median of all the differences between each local minimum and the higher of the two adjacent maxima (before and after the minimum). ``Depth" is defined by the amplitude (range) of the light curve within a single rotation. Note that this is different from the depth of a dip compared to the unspotted continuum (the true intensity deficit). It is more akin to what is measured by Kepler, which does not have the ability to measure the unspotted continuum \citep{Bas18b}. The set of all depths is called the ``variability curve" and its range is the total variability. Figure \ref{fig:example_mets} shows an example of range, depth, coverage, and variability.

Coverage is defined as the total integrated light deficit within one rotation period due to spots over the visible portion of the star. When there are no spots present on the star the coverage is equal to 0.  The higher the coverage value is, the more the star is covered by starspots. Generally we use the median of a metric over all rotations as the single metric for a full light curve. 

We measure the median, minimum, and maximum values of coverage and variability in each light curve. The variability curve is also sampled for its own broad maxima and minima (hills and valleys) and those are counted and measured for timing and relative amplitude. These global variability features are measured after smoothing the total variability curve over five rotations. To further characterize a light curve, we determined the widths and numbers of ``big hills" and ``big valleys". In this case, ``big" refers to widths or separations that are greater than half of that for the largest hill or valley. The same is done for coverage. There is one big variability hill in Figure \ref{fig:example_mets} at rotation 20, and two coverage hills at rotations 10 and 22. Note that the coverage hill at 10 is near a big variability valley; that is an instance where there are a lot of spots present but less difference between hemispheres (and two light curve dips per rotation).

It has always been noticed that the number of dip(s) in a light curve over one rotation is often either one or two. The next metric we discuss measures to what extent a light curve has essentially only one minimum per rotation (single mode), or whether there are two or more (double mode). The single/double ratio (SDR) is defined by \citet{Bas18a} as the logarithm of the ratio of the total time spent in single mode compared to the total time in double mode. This is a property of the light curve that has been simplistically misinterpreted by almost all of the literature to date on starspot light curves. It has been referred to as ``one spot"  or ``two spot" manifestations, and researchers have taken that too literally. The presence of phase coherence in the dip(s) also gets interpreted as ``active longitudes", and the motion of the dips in phase gets ascribed to physical motions of spots on the stellar surface. In fact, the number of minima per rotation is set by the hemispheric asymmetries of the whole spot distribution, since the observed intensity actually is determined by the hemispheric intensity at each time, which can be produced by any number of spots. The light curve is affected by projected (not actual) spot areas, by what fraction of a rotation spots remain visible, and by the geometric distribution of spots, which can evolve due to both spot evolution and differential rotation. This topic is discussed in more detail in the next section. 

We use a slightly modified version of the original SDR here. Each light curve dip is evaluated for the durations in time between its local maxima; these durations are the quantities of interest. We first smooth each light curve by one-eighth of a period. The reason for this is partially observational and partially theoretical. The noiseless models show few durations less than 0.2 period (because the spatial resolution of the light curve is low) and such dips have very small depths. In the observations, most dips with such small durations and depths arise from noise. Our choice of smoothing gets rid of most this problem without much loss in the information we are after. 

\begin{figure}[H]
    \includegraphics[width=18cm, height=10cm]{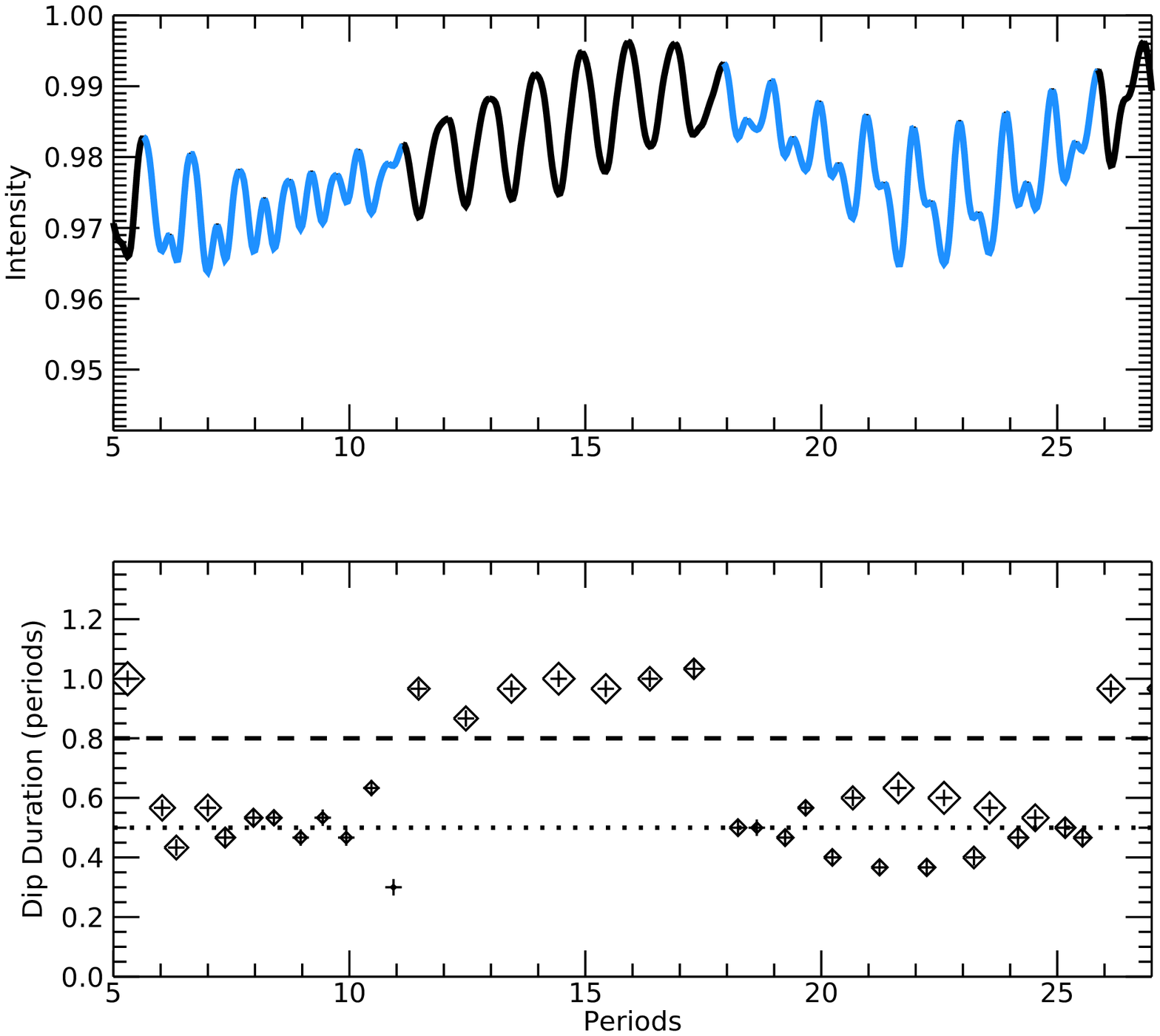}
    \includegraphics[width=18cm, height=6cm]{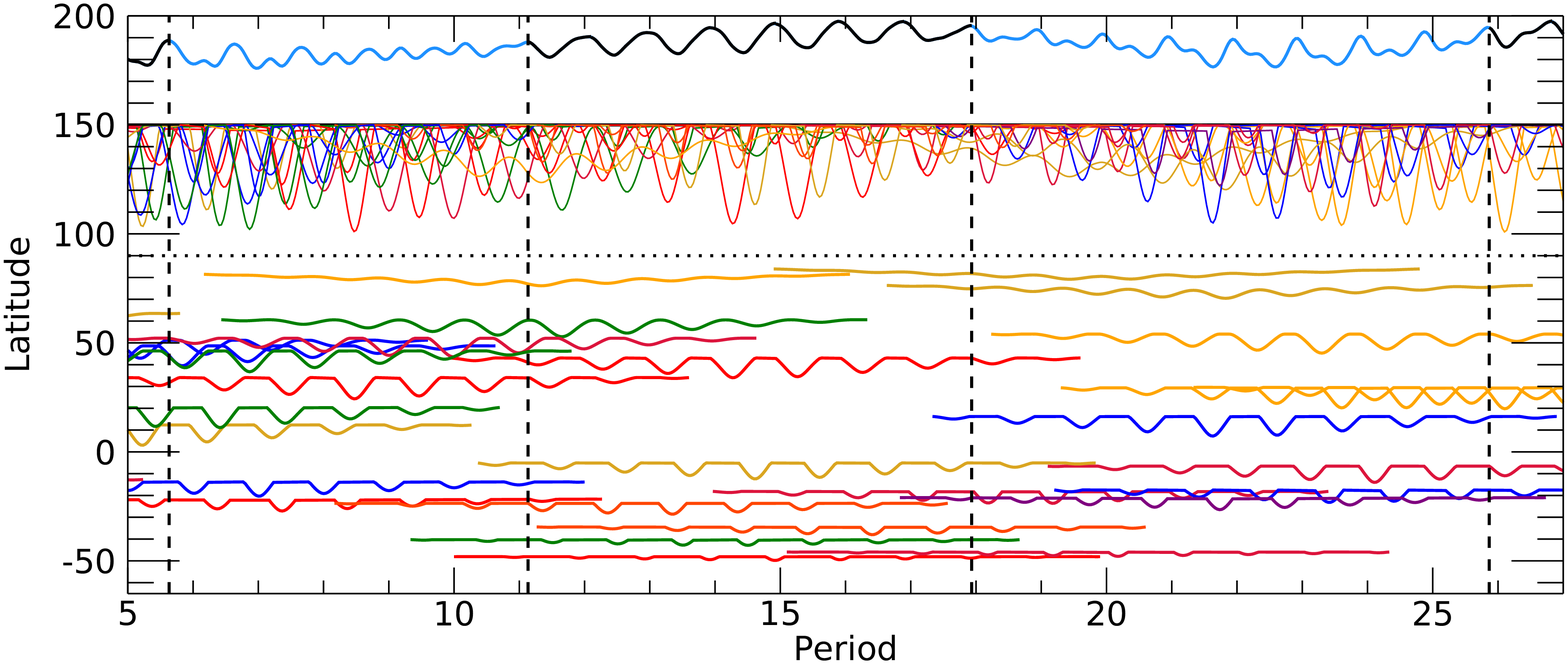}
    \caption{ Single/Double light curve segments. The top panel shows the light curve with single-dip segments (black) and double-dip segments (blue). The middle panel shows the dip durations in the top curve. If the value of the separation is greater/lesser than 0.8 of a period, we consider the segment to be  ``single-dip"/``double-dip". The size of each diamond is determined by the depth of the corresponding dip. The bottom panel shows the model spots that produce this light curve (as in Fig. \ref{fig:SameStarDiffIncl}). This model has spot number 12, lifetime 10, $i=60^\circ $, and no shear.}
    \label{fig:dbsnwid}
\end{figure}

If the value of the duration is greater than 0.8 of a period, we consider that dip to be single. \citet{Bas18a} found that there is a minimum in the distribution of dip durations at about that value, and that is true for the models as well. Double dips tend to be shorter (distributed about 0.5) and single dips tend to be longer (distributed about 1.0). If the duration is between 0.2 and 0.8, we label that dip to be ``double" (although there can be more than two such dips per rotation). Fig. \ref{fig:dbsnwid} shows an example of a light curve with both single- and double-dip segments. Such light curves are also seen in real data from Kepler, eg.  \citep{Bas18b} and other sources. For convenience, light curves that are entirely single-dipped are assigned an SDR of 2 and those that are entirely double-dipped are given a value of -2. 

\citet{Bas18a} showed that the SDR is strongly correlated with the rotation periods of the stars for which Kepler provides a secure period, and that these correlations also depend on the stellar temperature. One purpose of this paper is to further investigate what determines the SDR in different spot models. Notice that the ``oval" pattern of points in the middle panel of Fig. \ref{fig:dbsnwid} between rotations 20-25 is a signal of the presence of a ``drifting dip" in the light curve. The small dip is primarily to the left of the large dip at first but ends up on the right. This type of light curve behavior is often interpreted as a drift of the spot pattern due to differential rotation, but here there is no shear and the pattern arises simply from spots growing and decaying on different parts of the star.

One purpose of this paper is to render the construct of ``one or two spots" as antiquated. It is clear that the presence of one light curve dip per rotation is simply a result of the fact that a complex distribution of spots will generally leave one hemisphere of the star a little darker than its opposite (the bottom of the single dip in the light curve occurs at the time when the star looks darkest). The amplitude of the dip does not express the amount of spot coverage, but the difference between the hemispheres. The two sides of the dip need not be symmetric (and usually are not), and the smoothness of the dip is simply a manifestation of the fact that the light curve at any given time samples the full visible hemisphere. 

The presence of two dips per rotation requires a sufficiently complex spot distribution, but complexity does not imply a large number of spots. Instead it means a phase distribution that allows the light curve to rise and fall more than once during a rotation. Ironically, the presence of a larger number of spots can sometimes favor a single dip (the ``one spot" interpretation) because the aggregate phase separation pattern needed to produce two dips can be more difficult to attain with more spots. On the other hand, a double dip (``two spot" interpretation) becomes more likely as the spot lifetime gets short (2 or less rotations) because spots can significantly change size during one rotation, changing the asymmetry and causing a reversal in the light curve.

\begin{figure}[H]
  \includegraphics[width=\linewidth, height=11cm]{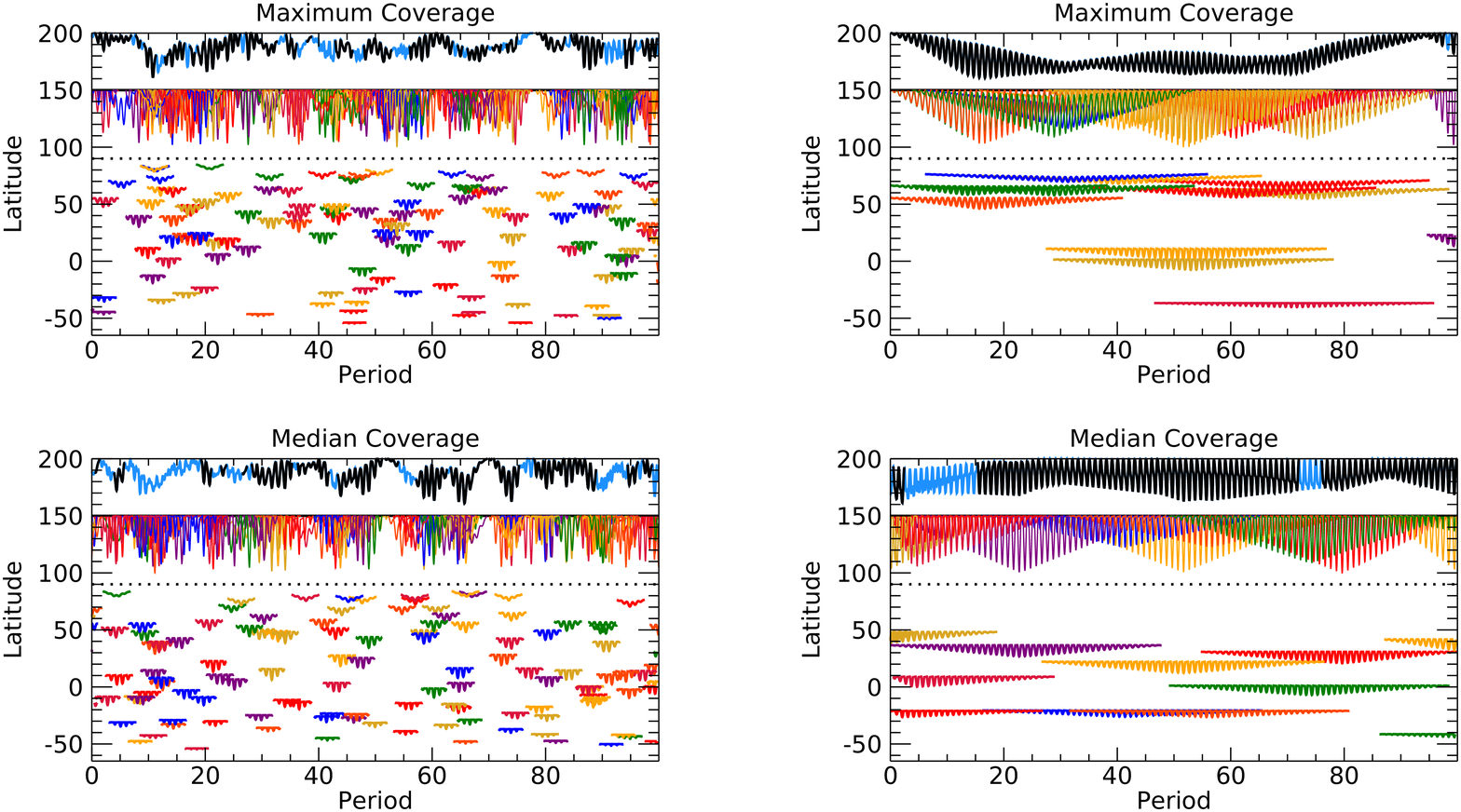}
\caption{Examples of short and long lifetime canonical models with no differential rotation. The panels on the left side are models with a spot lifetime of 5 rotations while the panels on the right side are models with a spot lifetime of 50 rotations. The trials shown are at high and median levels of coverage. }
\label{fig:SpotcloudDiffEv}
\end{figure}

Fig. \ref{fig:SpotcloudDiffEv} provides different examples of how spot behaviors affect light curves. It presents two different spot lifetimes for our canonical model. On the right are very long-lived spots with a lifetime of half the total observation period, or 50 rotations (so only a total of 12 spots are needed in the run). The upper right plot is for a case where the median integrated spot coverage per light curve is maximal for the 3000 cases with this set of parameters. Because the spots are so long-lived, the segments where the light curve remains single or double tend to last longer. The top right panel is virtually entirely single-dipped despite spots coming and going at various locations and times. Their random phase differences happen to always lead to simply a darker and less dark hemisphere. The high coverage is primarily due to the fact that the spots have latitudes closer to the sub-observer point. The lower right panel has a typical level of coverage. This particular case has more spots in the southern hemisphere which contribute less due to their diminished projected area. The spots in the segment from 5-20 are out of phase with each other enough to lead to double-dip rotations, and the dip drifts in phase as the relative strengths of the spots trade places. 

The left panels shows similar cases for spot lifetime 5; this requires 120 spots for spot number 6.  The light curve switches relatively frequently between single and double modes. The coverage doesn't change as much between its maximum and median values compared with the long-lived case. We see a complicated interplay of spots that leads to both single and double segments at a variety of coverage levels, and there can be rather short segments (usually double). There is a noticeably larger variability in the single versus the double segments (because of the increased hemispheric asymmetry, not because of the number of spots present). Some single and double segments have very similar coverage, but different variability. This is a contingent effect that depends on which spots appear and disappear, exactly when they do it, and where they are placed (since there is no shear). The inadequacy of a ``one or two spot" description of what is happening is made manifest by these examples (and across all the models). 

\section{Results \label{sec:Results}}  

We now discuss the behaviors of the various light curve metrics as a function of the model parameters. The point is to understand to what extent each of the metrics provides information about the underlying behavior of the parameters of interest, particularly spot lifetime and differential rotation. We also want to understand in what ways we can be fooled about the underlying meaning of variability changes.

\subsection{SDR \label{sec:SDR}}

\begin{figure}[H]
\centering                
\includegraphics[width=\linewidth, height=8cm]{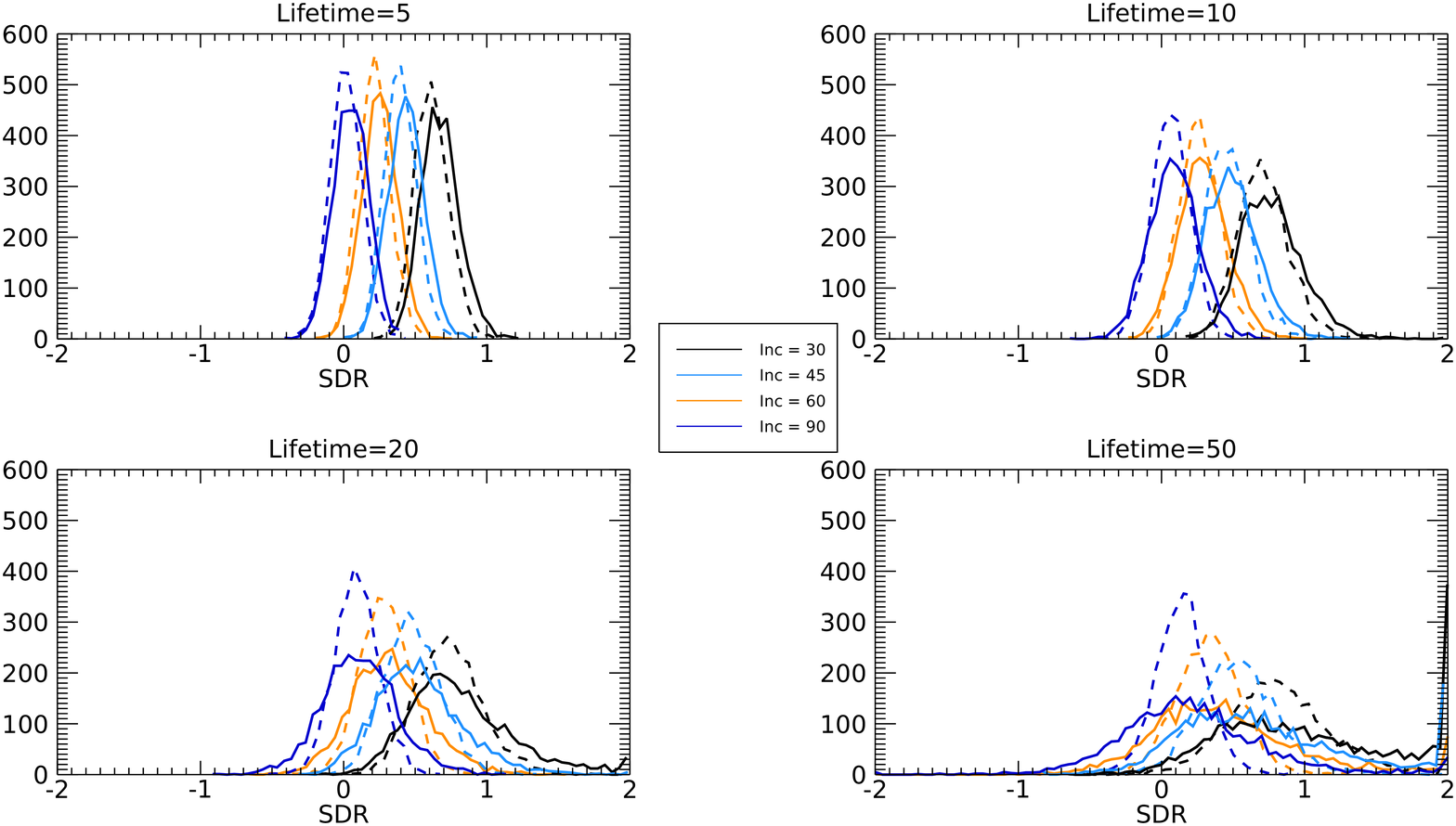}
\includegraphics[width=\linewidth, height=8cm]{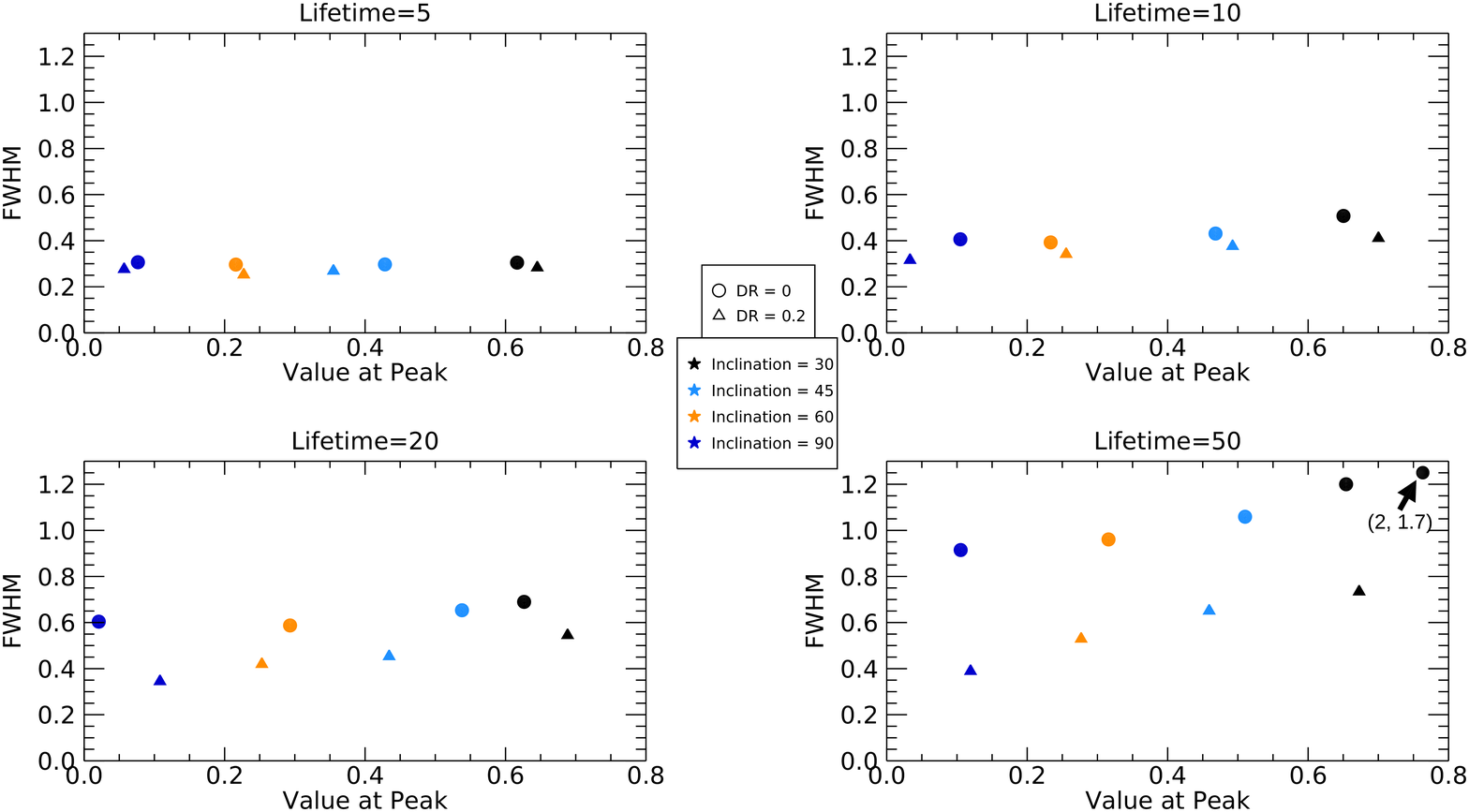}
\caption{SDR vs Inclination, Lifetime, and Differential Rotation. The top four panels show histograms of the (logarithmic) SDR for four different evolution cases (lifetimes of 5, 10, 20 and 50 rotations) each containing four different inclination cases (30, 45, 60, and 90 degrees by color). The solid lines show cases with no differential rotation, and the dashed line cases have a shear of 0.2 (the solar value). The bottom four panels show the FWHM vs. the SDR at the peaks of the distributions (all positive). Circles are cases without shear and triangles have solar shear. This scheme is repeated in all similar figures. }
\label{fig:SDRdiffincl}
\end{figure}

Figure \ref{fig:SDRdiffincl} demonstrates how the SDR changes with inclination, starspot lifetime, and the differential rotation of the star. It shows that as inclination decreases, regardless of spot lifetime or differential rotation, the SDR distribution moves to more positive values. This means that the light curves have increasingly more single-dipped segments at lower inclinations. We only show the results for solar differential rotation or none, but the same qualitative effects occur at twice solar shear. The same starting spot distributions are used with and without differential rotation, although of course the spots end up in different places at later times as differential rotation operates. The SDR is larger (more single) at lower inclinations partly due to the fact that (higher latitude) spots remain more visible throughout a rotation as the inclination decreases. Conversely, lower latitude spots disappear more often at higher inclinations, giving them a better chance to produce a double dip. 

A demonstration of this effect is visible in the top two panels of Fig. \ref{fig:SameStarDiffIncl} which have no shear. The light curve is single-dipped between rotations 60-80 for the 30$^\circ $ case, while it is predominantly double-dipped for the same segment when viewed from the equator. This is due to the increased influence of the red low latitude spot near rotation 65 and the blue and yellow ones near 80 and 85. On the other hand, the southern blue spot near 75 (which has a very low projected area in the low inclination case) does not cause a double dip simply because its phasing matches well enough with the high latitude spots. This reveals an unfortunate truth about the SDR: it is very contingent on the (random) phases at which spots appear. Differences in the longitudes of individual spots make the difference between a single- or double-dipped rotation. The SDR does contain some statistical information about spot distributions if measured over a large number of rotations, but is much less meaningful within a few rotations. The observational implication is that one cannot make a very informed statement about a star's spots if it is only observed for a few rotations.

The bottom two panels of Fig. \ref{fig:SameStarDiffIncl} suggest another promising aspect of the SDR, as they show it decreasing with increasing differential shear on the star. The star starts off with the same spot distribution in all the panels of the figure, but they begin drifting with respect to each other in the bottom two panels. It is hard to discern the drift except in the overlapping individual light curves, but the segment between 15-25 (for example) becomes double in the highest shear case while being single in the other three. Of course, any such effect depends on having spots at sufficiently different latitudes (where the shear is different). This would be a promising way to measure differential rotation if it were not for the fact that the spot lifetime also affects the SDR as discussed below; independent knowledge of the spot lifetime is required to break that degeneracy. 

The bottom panels of Fig. \ref{fig:SDRdiffincl} show that the most common values of SDR don't change much for short spot lifetimes with differential rotation compared to none. At longer lifetimes differential rotation tends to make the SDR less positive for higher inclinations. The peak locations of the SDR distribution for a given inclination, regardless of differential rotation or evolution, have similar values for the different lifetimes, with the exception of the lifetime 50 case. That case has a second peak at SDR of 2.0 reflecting cases that are entirely single. This can more easily happen when the spots last a long time and start off with the right spatial distribution. These results imply that the parameter that best spreads the SDR distribution for a particular activity level (or spot number in this case) is the stellar inclination, which is an observational rather than intrinsic property. 

As inclination decreases (for lifetimes greater than 5 rotations), regardless of differential rotation, the full width half max (FWHM) of the distributions increase. This means that there is a wider range of SDR values displayed by the light curves. The effect is compounded as lifetime increases - the FWHM of each case increases as lifetime increases. The case with the highest FWHM and peak value is that which has an inclination of 30 degrees, a lifetime of 50 rotations, and no differential rotation. This case resembles those thought to exist on very young stars which seem to have persistent polar spots. Adding differential rotation to a particular case consistently reduces the FWHM for longer lifetimes. The (upper) histogram panels show that the SDR distribution can reach more negative values without differential rotation than with it.

\begin{figure}[H]
\minipage{0.5\textwidth}
  \includegraphics[width=\linewidth]{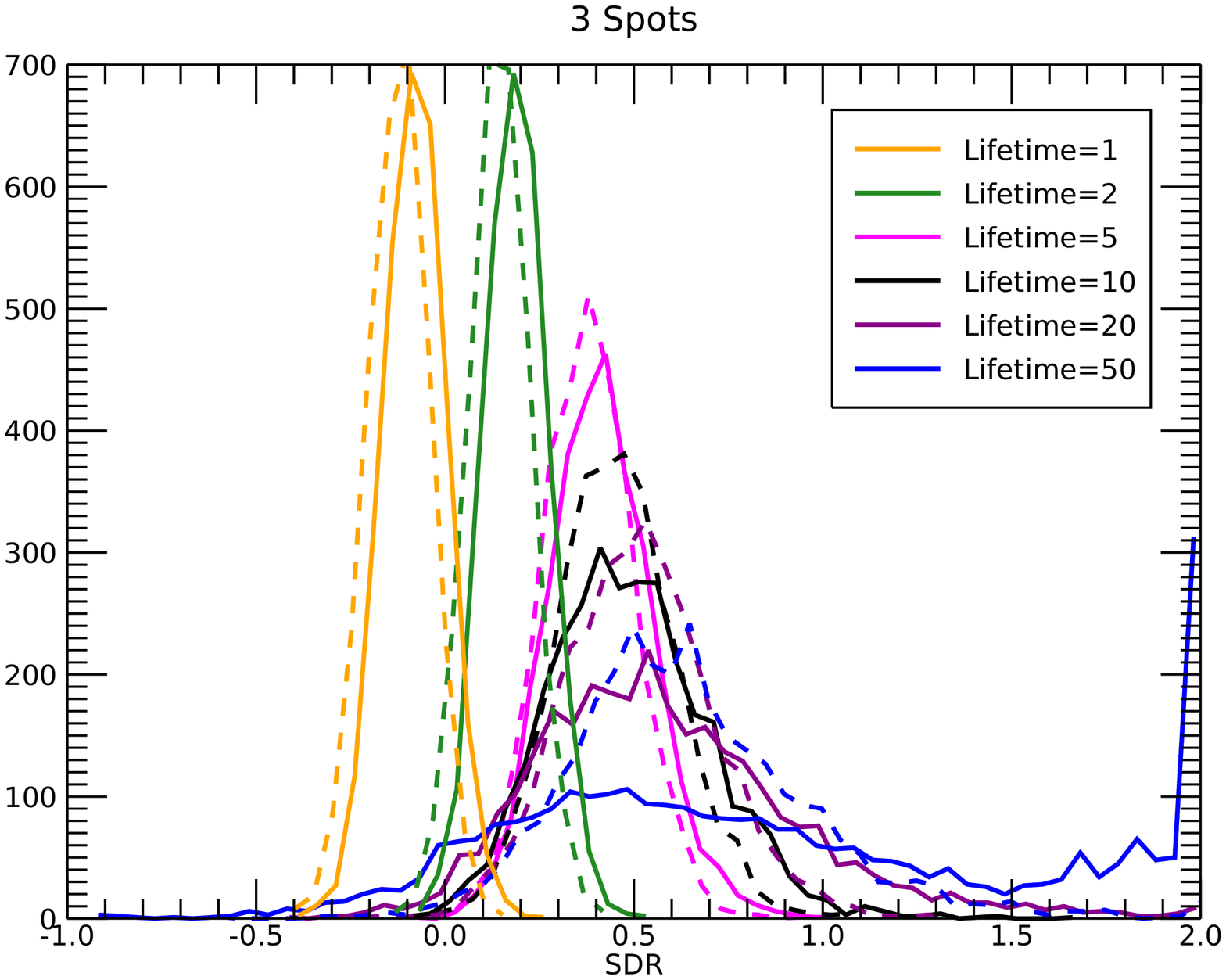}
\endminipage\hfill
\minipage{0.5\textwidth}
  \includegraphics[width=\linewidth]{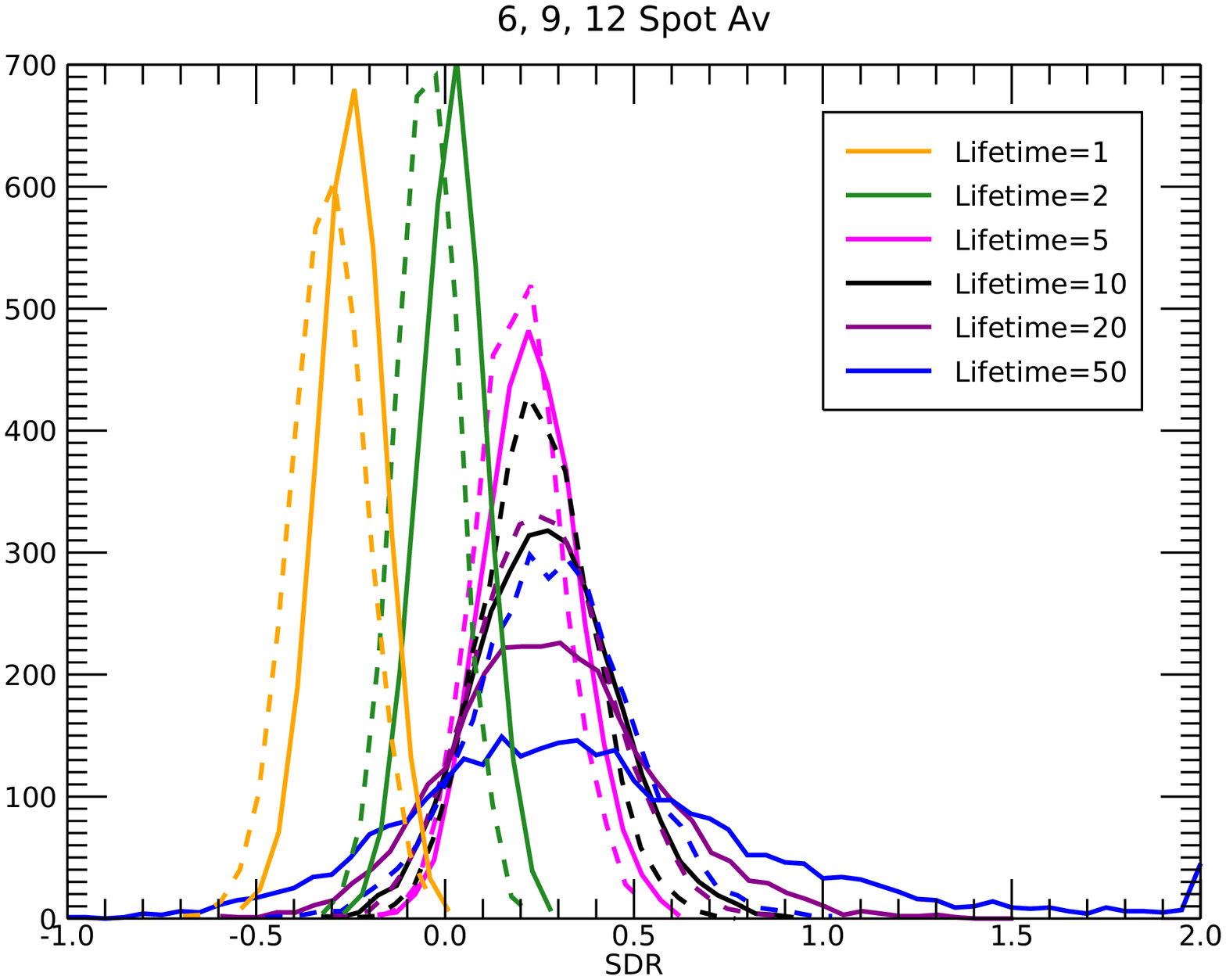}
\endminipage\hfill
\caption{SDR vs Spot Number and Evolution. The distributions of SDR for two different spot number cases and a variety of spot lifetimes ($i=60^\circ $). The left panel is for 3 spots, and the right panel is for the average histograms of 6, 9, and 12 spots (those cases all look very similar); dashed lines indicate solar shear. In general, the SDR becomes more positive as the lifetime increases. For very long lifetimes it spreads due to the small total number of spots that remain essentially in fixed positions when there is no shear. }
\label{fig:SDRdiffspots}
\end{figure}

Figure \ref{fig:SDRdiffspots} shows the distributions of SDR for all the various lifetimes when the spot number is only 3, and when it is larger. We make this division in spot number because the behavior of the solutions are very similar for spot numbers 6 and larger. The SDR is almost entirely positive (more single-dipped than double-dipped) for lifetimes between 5 and 20 rotations, slightly more so for 3 spots than for larger spot numbers. The lifetime 2 case is also positive for 3 spots, but almost symmetric about zero for more than 3 spots, extending to -0.3. Lifetime 1 is the only case that is primarily negative (mostly double-dipped); the number of spots does not matter much in this case but having more spots pushes it a little more negative. In this most solar-like case, the complexity of the spot distribution can change significantly during half a rotation, which makes it more likely that the light curve ends up double-dipped. 

At a spot lifetime of 50 rotations, there is an additional peak at the purely single-dipped end of the graph (+2.0) for 3 spots. This reflects the fact that the small number of long-lived spots involved in that case can sometimes manage to all stay in a sufficiently asymmetric longitudinal configuration for the whole run. That is the underlying reason that the distributions are displaced a little toward positive SDR for all the 3 spot cases: it is statistically easier to be in a single-dip rather than double-dip configuration with fewer average spots, since the latter requires a particular type of complexity. The lifetime 50 case also displays a larger FWHM for the SDR distributions, extending nearly to -1 (more double-dipped) for all the spot numbers. This is also due to the smaller total number of spots present throughout a run, which can maintain a more double-dipped configuration for longer if they happen to be placed that way. Differential rotation did not make a significant difference in almost all cases, the exception being lifetime 50. For very long-lived (or permanent) spots, the presence of differential rotation constrains the SDR to being similar to lifetime 20 results. This is because the exceptional cases that lead to the broad wings with no shear are not able to be maintained for the whole run when shear is present.

Spot numbers from 6-12 all produced similar SDR distributions; this is likely to continue for even larger spot numbers. Real stars more active than the active Sun lie in this regime, which should lead to essentially positive SDRs for them almost all the time. That is not what is seen in the observational data \citep{Bas18a}, which is a puzzle worth further investigation. Our preliminary explanation is that the observations can be contaminated by noise (the models are noiseless). This noise will have the effect, if sufficiently comparable to the intrinsic starspot variability, of introducing additional small dips into the light curve and thus producing more apparently double-dip segments. This is obviously more of a problem with less active (more slowly rotating) stars, which have smaller variability metrics. One test of this could be made by examining how the SDR varies with apparent magnitude for otherwise similar stars.

\subsection{Coverage, Range, and Variability \label{sec:Coverage}}

\begin{figure}[H]
\includegraphics[width=\linewidth, height=7.5cm]{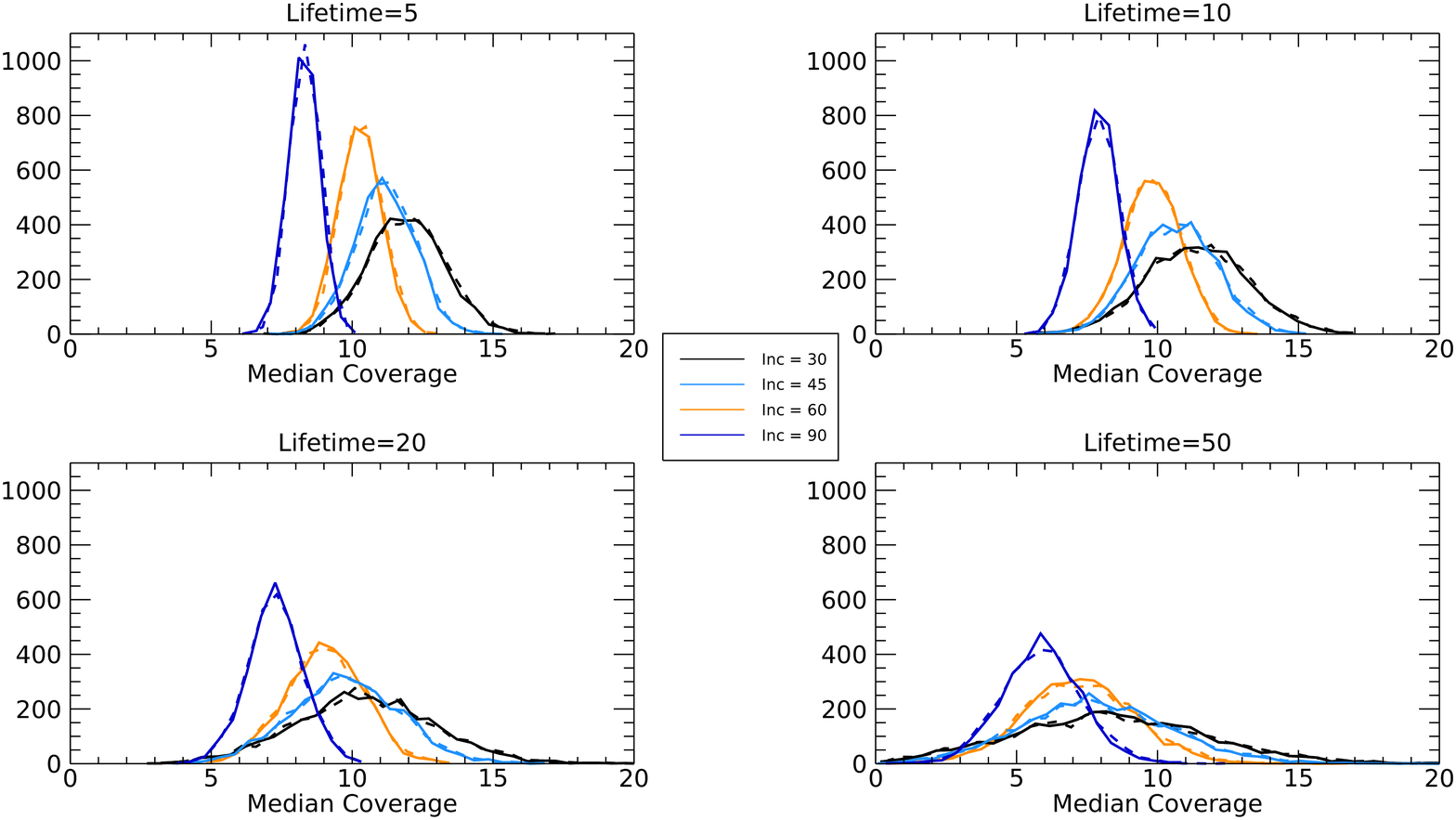}
\includegraphics[width=\linewidth, height=7.5cm]{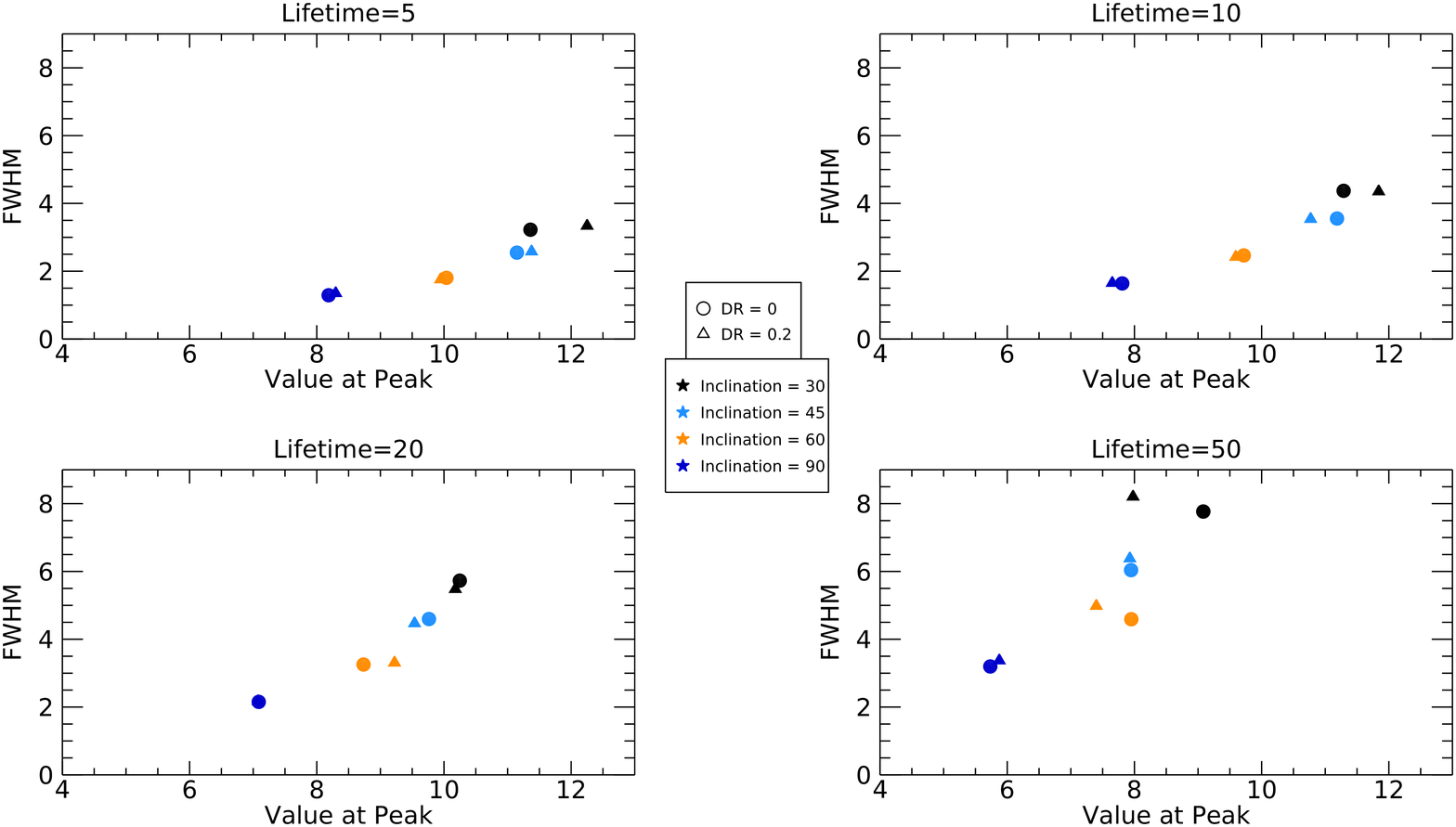}
\caption{Coverage vs Inclination and Lifetime. The panel above shows histograms of the median coverage (in ppt) for the same cases as in Figure \ref{fig:SDRdiffincl} with and without solar differential rotation (which makes little difference, as it should). }
\label{fig:Covdiffincl}
\end{figure}

Figure \ref{fig:Covdiffincl} shows how spot coverage is affected in our various test cases. What is shown is the distribution over all light curves of the median coverage for each light curve. Differential rotation makes no difference in coverage, because that metric is designed to capture the spot coverage over the visible star during a whole rotation. Even if spots move around, they still contribute to it (except those that cross the zero phase boundary). For a given inclination, as spot lifetime increases the peaks of the distributions shift towards smaller coverage and the distribution of median coverages widens. These trends are both due to the fact that for a fixed spot number (6 in these cases), the total number of spots over the run decreases as the spot lifetime gets longer. As illustrated in Fig. \ref{fig:SpotcloudDiffEv}, the overall behavior of the light curve is determined by about 12 spots in the case of lifetime 50, while 120 spots are involved in the light curve for lifetime 5. Once a long-lived spot appears, it spends half of its time at less than half its maximum size. The random birthdates and positions for the short-lived case lead to a more variable light curve because there is more chance for several spots to overlap strongly. There is also a larger likelihood of several spots being present near their maximum size. This makes the median  coverage (at a given inclination) larger in the short-lived case than in the long-lived case. However, there is a larger dispersion in median coverages for longer spot lifetimes because there are fewer chances to sample a large set of possible overlaps, which means that the individual light curves differ overall from each other more strikingly. The trends with lifetime are therefore partly an effect of how we construct our models, so they must be used to interpret real stellar data with some caution.

For a given spot lifetime, as inclination decreases the overall spot coverage on the visible hemisphere increases. This is partly because fewer of the spots are hidden as the star rotates. It is also partly due to our scheme for placing spots, since we don't place them where they can never be seen. This has the effect of concentrating spots closer to the pole as the inclination is lowered, where the spots remain visible for more of a rotation.  Since we place spots at random visible locations in any given trial, larger inclinations allow for a larger total stellar area on which spots can be placed. This leads to a lower probability that several spots will be visible at the same time. When a greater number of spots are visible at the same time it leads to a stronger light deficit, and makes the median coverage increase. Thus lower inclinations tend to generate higher median coverages. The other effect is that the distribution of coverages broadens for lower inclinations at a given lifetime.  This is due to the fact that at lower inclinations once spots appear they tend to remain in sight. When a higher than average number of them happen to be on the star the net effect is enhanced despite their relative phases. By contrast, when the same situation arises at high inclinations, the spots can be out of phase with each other and be hidden, so they do not have the same cumulative effect. This not only lowers the median coverage but it reduces the extreme values that it can attain.

\begin{figure}[H]
\centering
     \includegraphics[width=\linewidth, height=11cm]{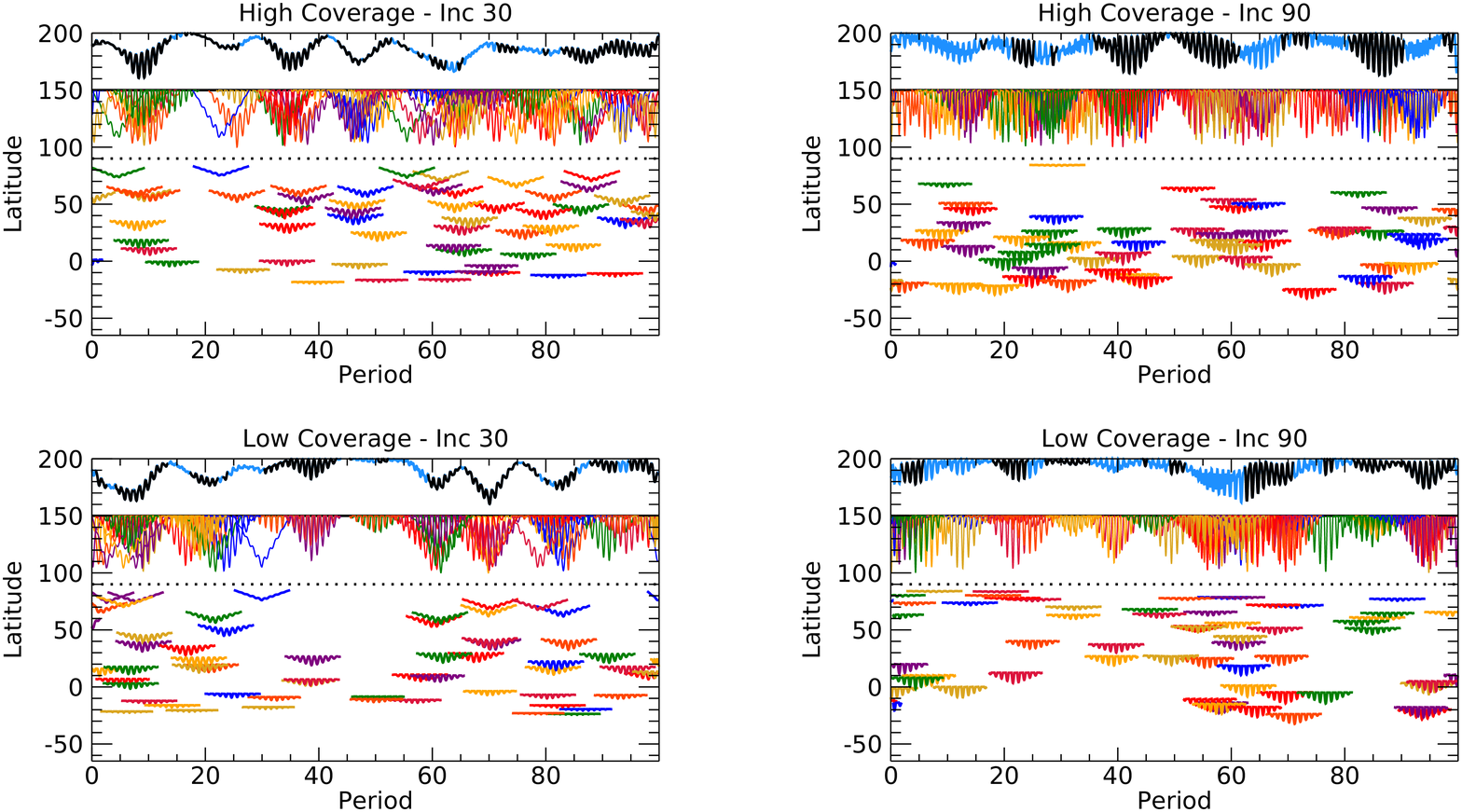}
\caption{ Inclination effects on light curves. This is similar to Figure \ref{fig:SameStarDiffIncl}, with 2 panels each for trials at high and low coverage values. These models have spot number 6, lifetime 10 rotations, no shear, and $i=30^\circ $. The left panels are viewed from the initial inclination while the right panels contain similarly chosen coverage examples when observed from 90$^\circ $. }
\label{fig:Base30Spotcloud}
\end{figure}

Examples of these effects are shown in Figure \ref{fig:Base30Spotcloud}. The left panels display two cases where spots are generated for a star at $i=30^\circ $ with lifetime 10 rotations, spot number 6, and no differential rotation. The cases are chosen for high and low median coverage levels, though of course the instantaneous coverage varies throughout each one.  On the right are shown a similar set of cases for a star at $i=90^\circ $. It is immediately apparent that the light curves exhibit more double-dipping (have lower SDRs) at the higher inclination, as discussed above. What might be over-interpreted as ``activity cycles" in the left panels are just random bunchings of spot birthdates. The variability curves have 3 or 4 ``big"  valleys; equivalently they occur for every 2-3 spot lifetimes. 

In general these examples show that contemporaneous high latitude spots produce stronger integrated deficits, not just because the spots have a larger projected area but because they are always visible. The top left panel from rotations 55-70 or the bottom left panel from 0-15 show stronger sustained deficits than the top right from 20-40 or bottom right from 50-70 even though all those segments have a lot of overlapping spots. The individual deficit curves for the high latitude spots on the left are ``V-shaped" because those spots are always visible as they grow and decay. The ones on the right return to a flat top (zero deficit) during each rotation because they disappear behind the star. The top left panel is a high coverage case because the (randomized) spot distribution is skewed towards high latitudes, so more spots are visible for longer times. Similarly the top right panel shows maximal coverage because the overall spot distribution is skewed towards the equator. 

Close examination of these examples reinforces the essential causes of single and double segments. In the upper right panel for example, the light curve is single between 50-65 and double between 65-70. Examination of the overlapping individual curves shows that similar numbers of spots were in phase with each other in the first part, but some new ones grew out of phase in the second part (due to their random birthdates and longitudes). In the lower right panel more or less the same thing happens in reverse between 55-70. Returning to the upper right panel, it is hard to see from looking at the spot map why a large amplitude single segment occurs from 80-90 surrounded by small amplitude double segments, but most observers would likely conclude that the star had become more active then (it is mostly due to the fact that the southern red and blue spots happen to be in phase). 

The variability in the double segments tends to be smaller than in the single segments regardless of the actual spot coverage, because spots in phase can produce a larger differential variation. This is a general relation; the median variability of single segments is generally about twice that of double segments. The fact that a light curve remains single for a while does not imply the presence of active longitudes. It just means that the spot distribution is asymmetric enough to produce one full darker hemisphere. These examples imply that it is difficult to infer clear and specific conclusions from the fact that the light curve happens to be in a single or double mode. Larger variability also does not necessarily imply higher spot coverage -- to correctly infer coverage levels requires knowledge of the placement of the light curve relative to the unspotted level (cf. \citet{Bas18b}). 

\begin{figure}[H]
\includegraphics[width=\linewidth]{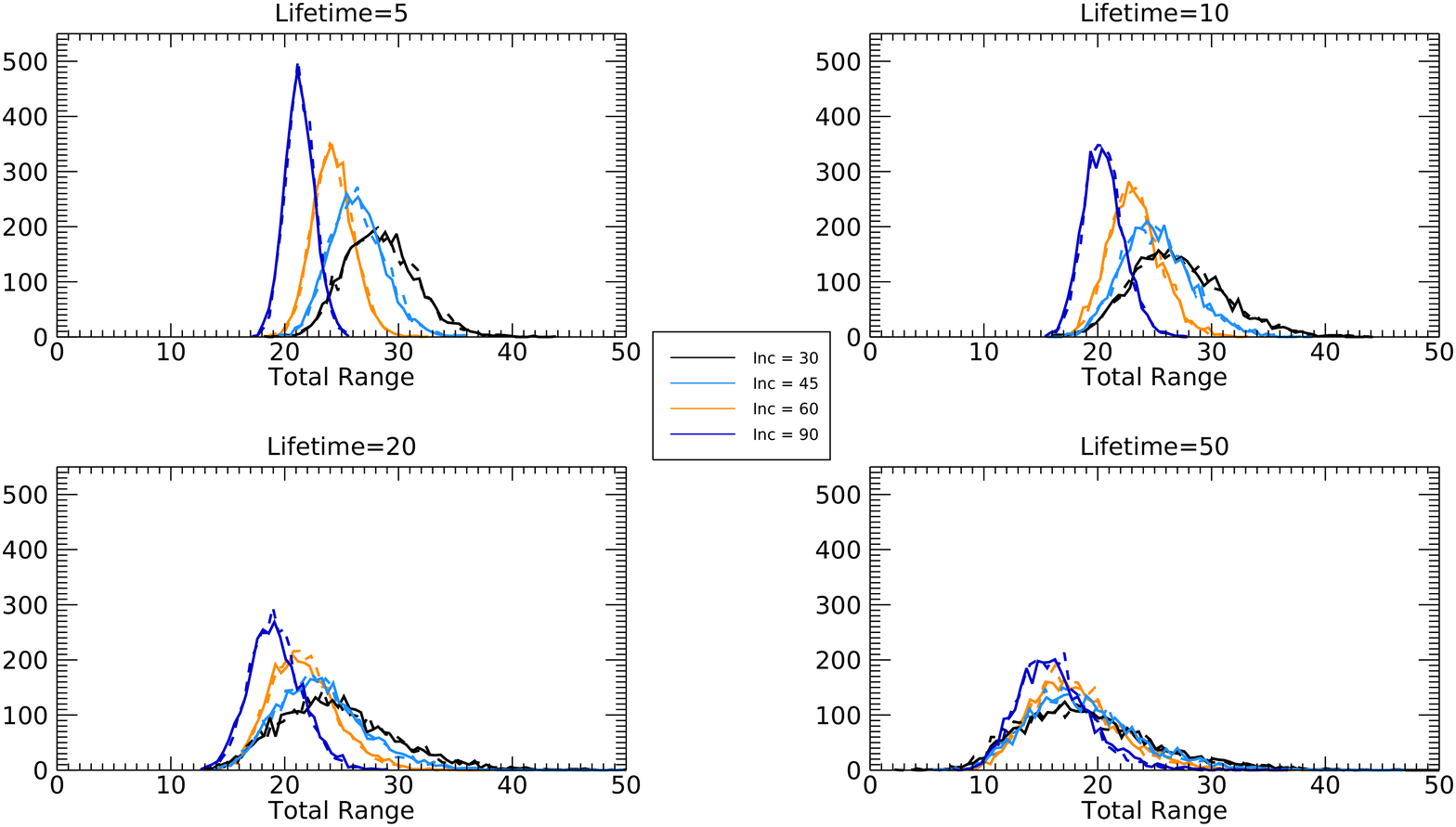}
\includegraphics[width=\linewidth]{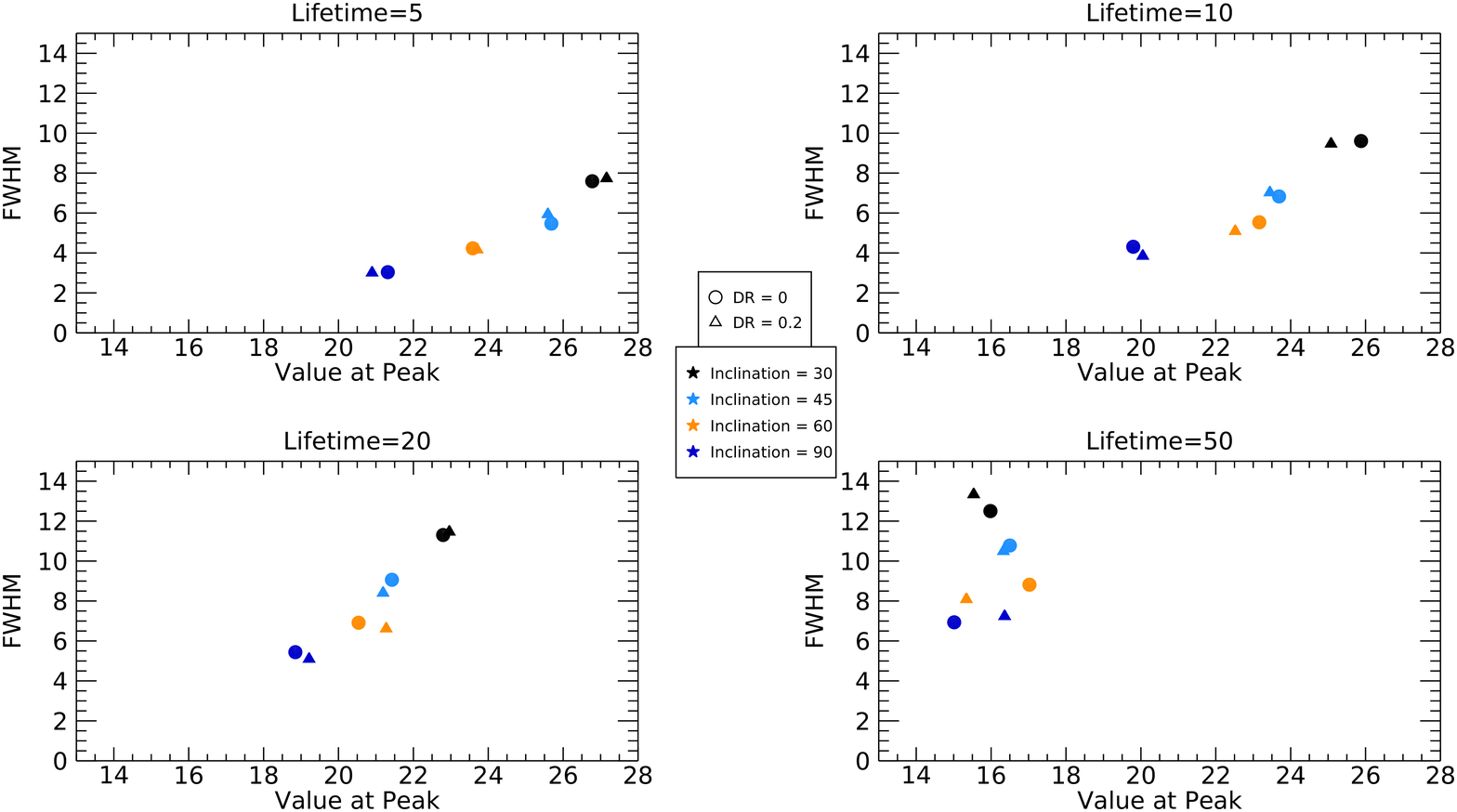}
\caption{Total Range vs Inclination and Lifetime. The panels above show 
the total range (in ppt) for the same cases as in Figure \ref{fig:SDRdiffincl}. The dashed lines in the left two plots represent the same cases as the solid lines except with a shear of 0.2. The peak values of the histograms increase with decreasing inclination except for the lifetime=50 case. Differential rotation makes little difference; none in the $i=90^\circ $ case. }
\label{fig:TRdiffincl}
\end{figure}

Figure \ref{fig:TRdiffincl} displays the behavior of the total range in a similar set of trials as discussed for coverage. For a given inclination, as spot lifetime decreases the peaks of the distributions shift towards larger ranges. This has the same cause as the similar behavior of the coverage, namely that for shorter lifetimes there are more opportunities for a larger number of contemporaneous spots. In general, the light curve is simply more variable when the lifetimes are shorter. For a given lifetime, as inclination increases the distribution of total ranges decreases, again for similar reasons as with coverage. The distribution of total ranges also shrinks with shorter lifetimes. This is a consequence of the fact that when spot lifetimes are long, the contingent appearance or decay of a given spot has a larger and longer lasting effect on the light curve, since it happens more infrequently. In addition, as lifetime increases and the distributions spread out, the effect of inclination on the total range is less distinct. Differential rotation appears to have very little effect on the total range of the whole light curve.

\begin{figure}[H]
\includegraphics[width=\linewidth]{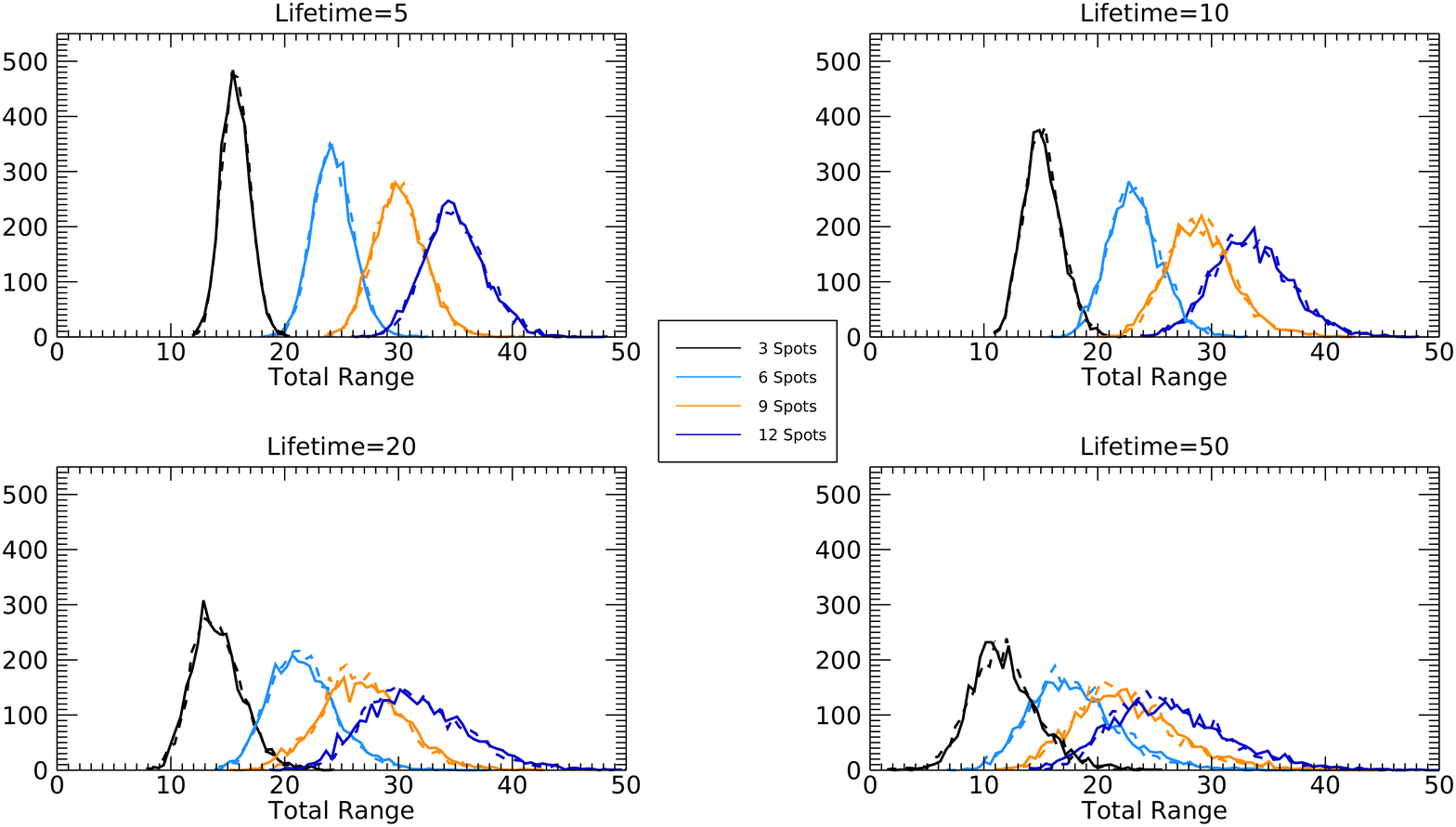}
\includegraphics[width=\linewidth]{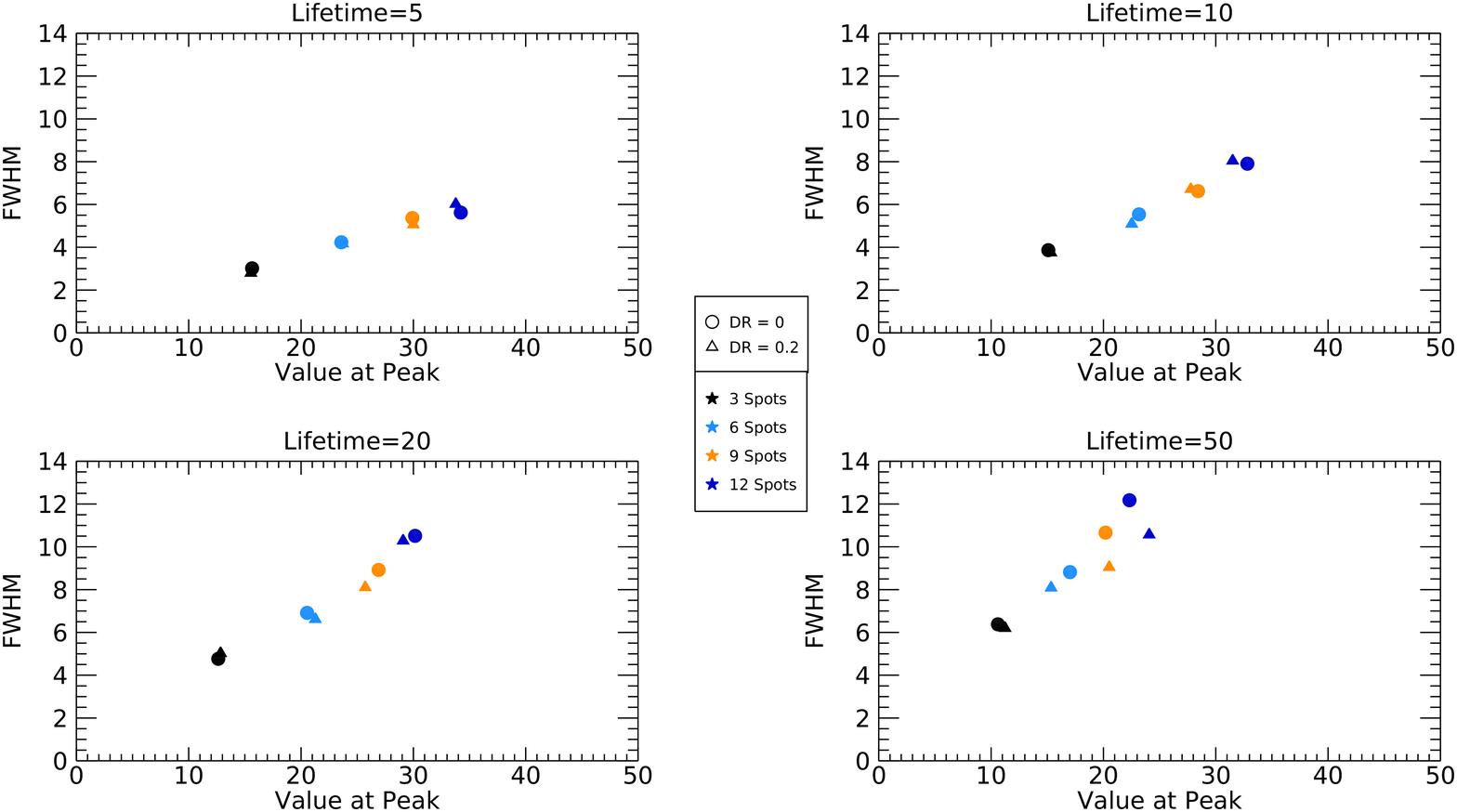}
\caption{Total Range vs Spot Number and Lifetime. The panel above shows the total range (in ppt) for four spot lifetimes. The range distribution both moves to higher values and spreads as spot number increases. Spot number is the dominant driver of the peak location; increasing lifetime increases the FWHM. Shear (dashed lines) is not important. }
\label{fig:TRdiffspots}
\end{figure}

Figure \ref{fig:TRdiffspots} displays the total range as spot number and spot lifetime change. As spot number increases, the total range increases regardless of spot lifetime. This is natural, since the range can become larger if more spots are available to contribute to the total light deficit (although there also have to be times where the deficit is minimal, since the range captures the difference between minimal and maximal deficits). The peak of the range distribution is fairly similar for a given spot number at various lifetimes, although it drifts to smaller ranges as the lifetime increases. The overall distribution also becomes more spread out (extending at both the low and high range ends) as lifetime increases. This is again due to the opportunity for more statistical variation as the total number of spots involved in the run decreases with increasing spot lifetime. Differential rotation again makes no difference to the total range for the various spot numbers.

The total range is responsive to coverage changes as well as spot distribution asymmetries. It is a metric that cannot be directly measured in the Kepler light curves, however, because the absolute stellar intensity level is not measured, and each quarter is reduced separately \citep{Bas18b}. The parsing of a model into ``quarters" would depend on how long the physical rotation period of a star is compared to the length of a quarter (90 days for Kepler). Rapid rotators might have 30 rotations per quarter, while a slow rotator might have only two or three. The median range of a light curve is certainly different when it is divided into chunks of 3 versus 30 rotations. 

\begin{figure}[H]
\includegraphics[width=\linewidth]{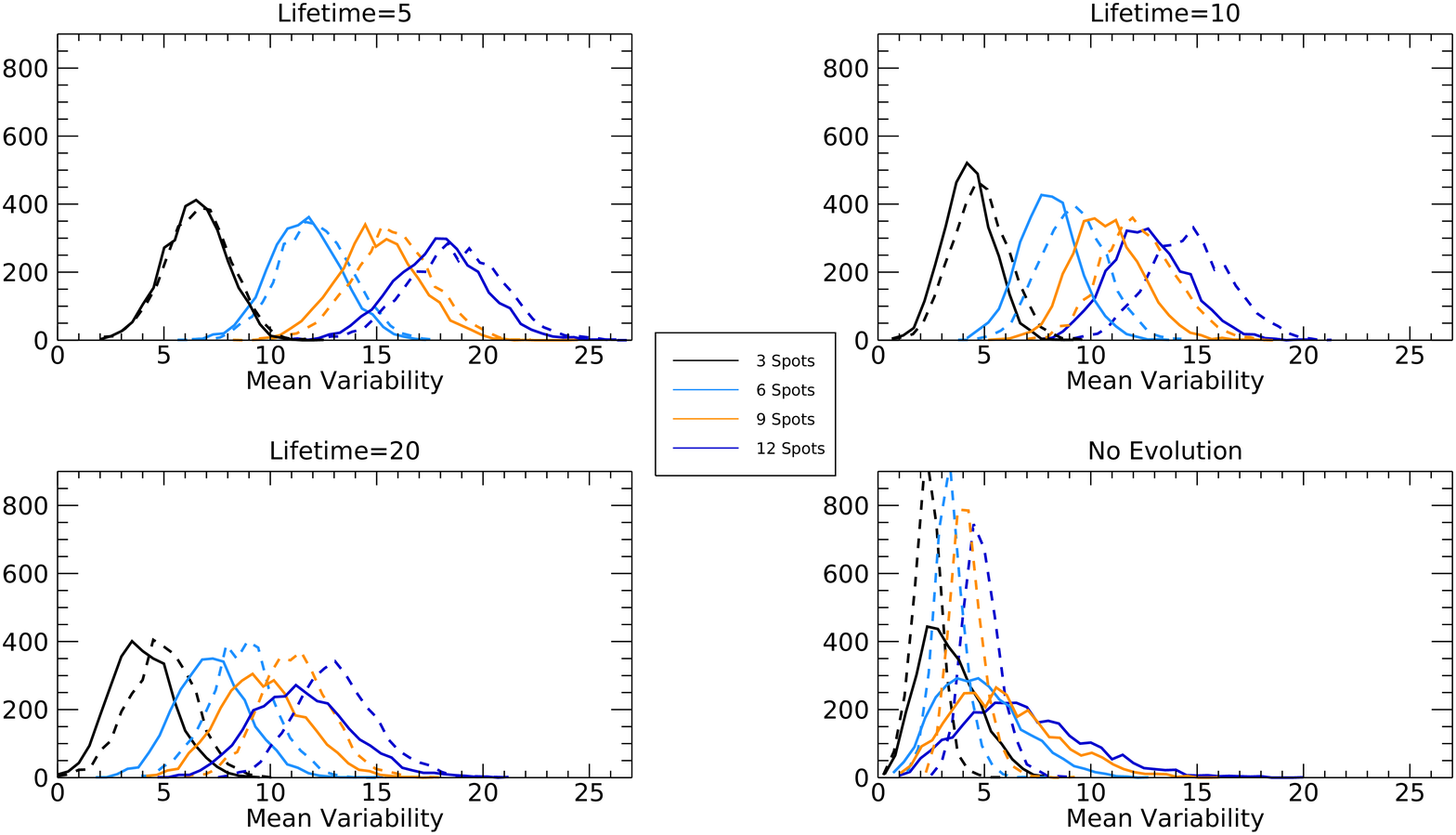}
\includegraphics[width=\linewidth]{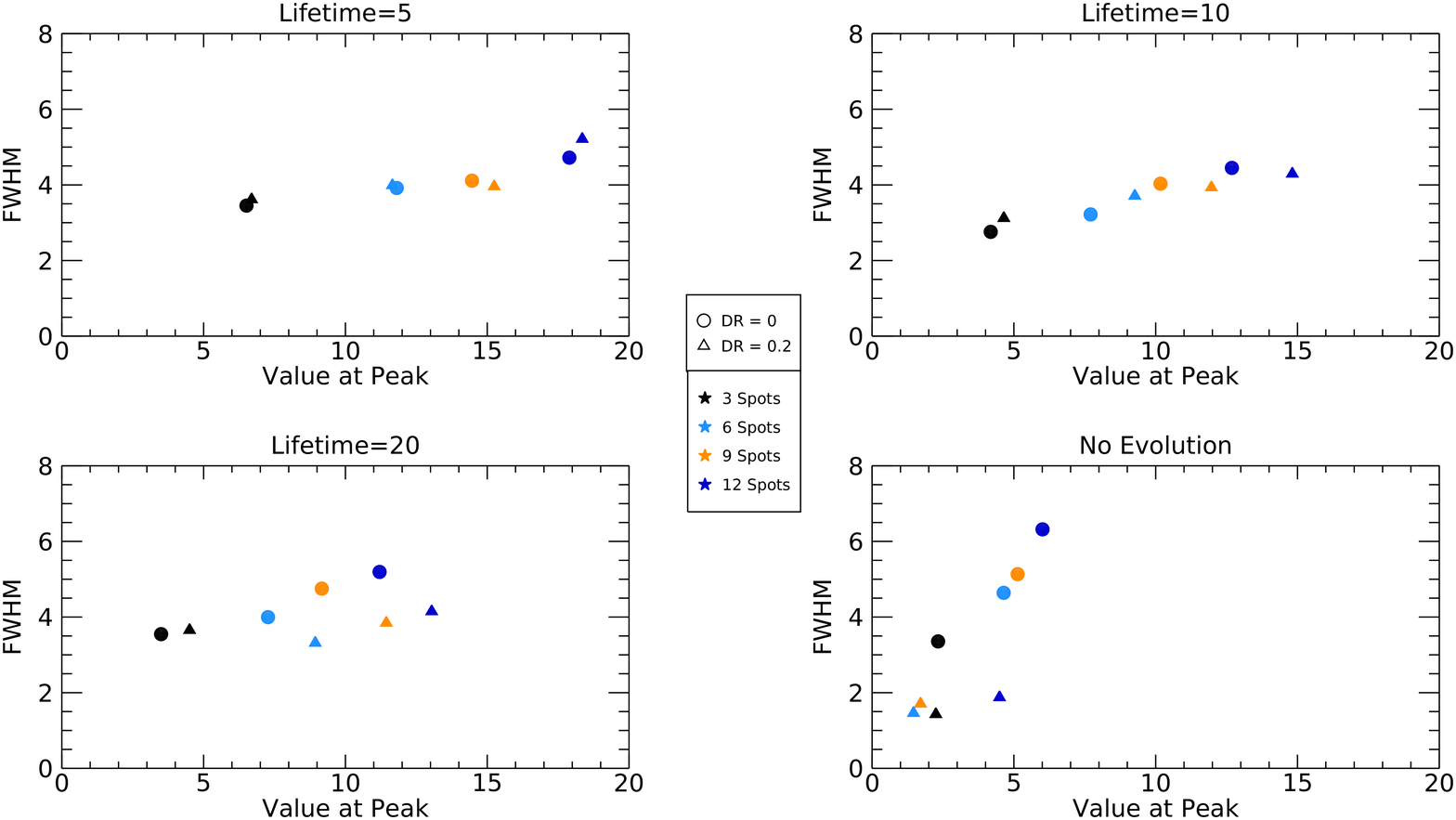}
\caption{Mean Variability vs Spot Number and Lifetime. The panel above shows the variability (in ppt) for four spot lifetimes. The variability distribution both moves to higher values and spreads as spot number increases. Shear (dashed lines) plays an important role in tightening the distributions for long spot lifetimes, but doesn't matter much for short lifetimes.  }
\label{fig:MeanVardiffspots}
\end{figure}

What can be measured in any case is the variability. Figure \ref{fig:MeanVardiffspots} shows how the mean variability behaves as a function of spot number and lifetime. We choose the mean rather than the median for this metric to better capture the extremes of how deep the dips get (the median is insensitive to that). The behavior of the mean variability is similar to that of total range (Fig. \ref{fig:TRdiffspots}) in that it increases with the spot number. This offers a possibility of being able to say something about the number of spots that are on a star, especially for small spot numbers. For short lifetimes differential rotation makes little difference for both metrics, but for longer lifetimes it begins to matter for mean variability. As the lifetime gets longer (10 rotations and above), its distributions for more than 6 spots increasingly overlap, although differential rotation keeps them sharper. 

We have already discussed the variability when talking about the amplitude of the differential light curve in single-dip and double-dip segments. These changes, along with changes in coverage that last for many rotations, are the drivers for the longer term changes in variability that we called variability or coverage ``hills" and ``valleys" in Section \ref{sec:Metrics}.  In general, as spot lifetime increases, the number of variation valleys over 100 rotations decreases from 12-15 for lifetime 5 to 3-4 for lifetime 50. That is not surprising, since the overall spot distribution changes more slowly at long lifetimes. Inclination doesn't have much of an effect on this metric. It is only significantly affected by differential rotation for long spot lifetimes, where twice solar shear causes up to twice as many variation hills. When the lifetime is 50 rotations there are also a significant number of cases where no big hills occur within all 100 rotations in a run because of the small number of total spots involved. 

Figure \ref{fig:MeanVardiffspots} shows the behavior of the variability with spot number and lifetime. It is qualitatively similar to the total range, and most of the discussion above explaining the trends also applies to the variability. This is the quantity more readily measured in Kepler data, however. It is instructive here to examine the case of no spot evolution, which exaggerates some underlying principles. This is displayed in place of the lifetime 50 case (lower right panel). Spot number now makes less of a difference regardless of shear. That is not surprising, inasmuch as with no evolution and no shear, whatever spot distribution occurs at the beginning simply repeats exactly every rotation. Essentially, those 3000 models sample a much smaller set of total configurations than models that have to keep replacing spots in random positions. What is interesting is the striking difference between the no shear case and that for (solar) shear. 

The variability is much more concentrated at low values with shear. The no shear cases with higher variability are ones with especially asymmetric initial distributions. The addition of shear has a qualitative effect similar to that of spot evolution -- it makes the variability higher as a function of spot number. But the peaks in the shear distributions are all within 2-5 ppt instead of 5-14 ppt for lifetime 20. This effect for variability is similar to what happens for total range (not shown), except that the no shear and shear cases share similar spans. Essentially the shear reduces the variability distribution peaks to values similar to those that obtain from restricting the spot number to 6 or less with no shear.

\subsection{Periodogram Analysis \label{sec:Periodograms}}

\begin{figure}[H]
\includegraphics[width=\linewidth]{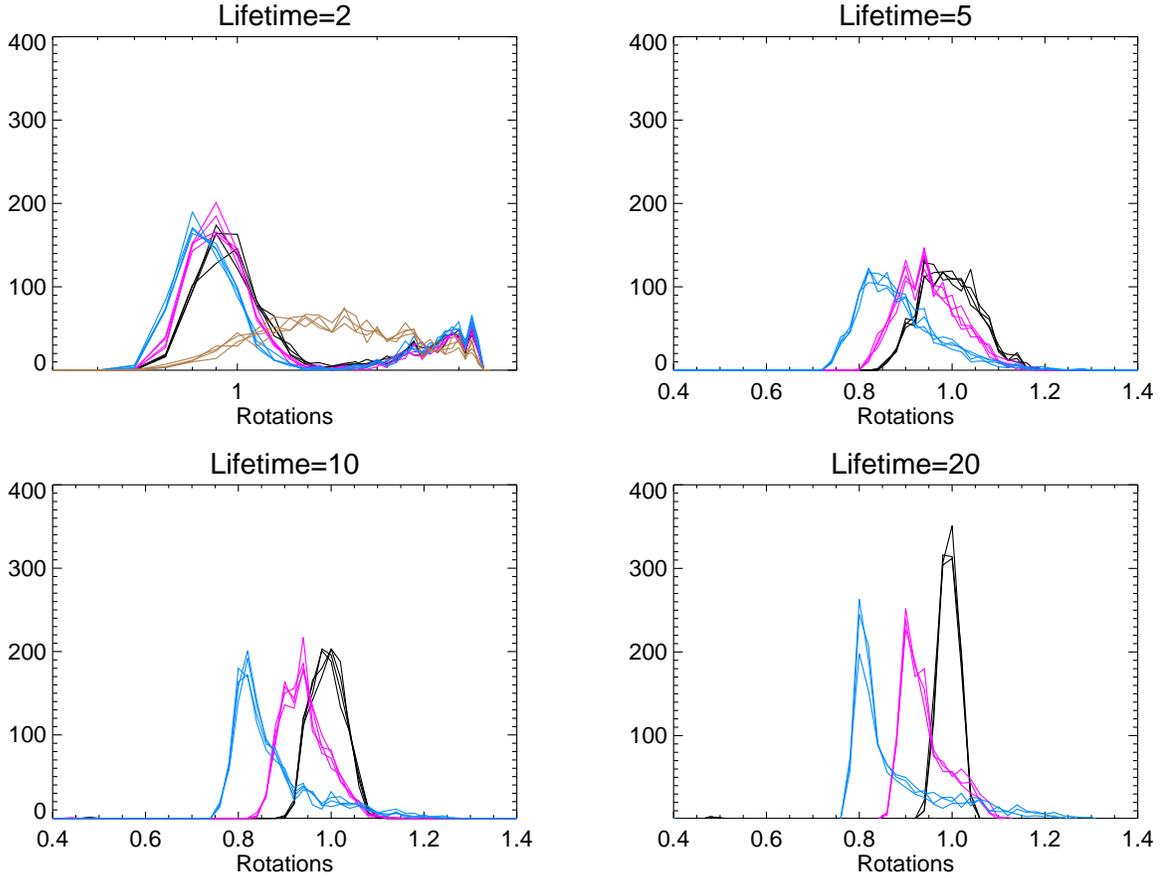}
\caption{ Periods derived from periodograms. Each panel contains models with the indicated spot lifetimes. The black curves are for all four spot numbers with no shear, the magenta curves are for solar shear, and the blue curves for twice solar shear. The true period in all cases is unity. In the top left panel the abscissa has a log scale, and the curves extend out to a period of 3.5. The brown curves in that panel are solutions for Lifetime=1 (see text).}
\label{fig:PdgProts}
\end{figure}

Another characteristic of this model set is that it provides a fertile ground for testing a common metric of starspot light curves that has been used to infer stellar rotation periods, and attempted as a diagnostic of differential rotation. That metric is the periodogram; here we employ the Lomb-Scargle version of it \citep{Zech09}. The rotation period is unity for all the models (for the ones with differential rotation that is the period at latitude 45$^\circ $); our time scale is expressed in fractions of a rotation period. It is a simple matter to apply a periodogram calculation to each model run and see how well it does at discovering the period, which we simply define as the location of the highest periodogram peak. Fig.  \ref{fig:PdgProts} shows the results for four spot lifetimes from 2-20 rotations and various spot numbers, all at inclination 60$^\circ $. For this analysis we analyzed light curves 50 rotations long in 1000 models for each parameter set. Three values of differential rotation are shown with the different colors (no shear is black).

The first result is that the longer the spot lifetime, the easier it is for the periodogram analysis to correctly detect the period. In fact, for spot lifetimes 10 rotations or longer with no differential rotation, the technique is accurate to within 5\%. The standard deviation of period determinations is somewhat worse at lifetime 5, closer to 10\%. Adding solar differential rotation lowers the typical measured period to a little over 0.9 rather than unity. Twice solar shear lowered it to 0.8; apparently the faster moving spots are favored, perhaps because they are seen again before a rigid rotation would be over. The distribution of highest periodogram peaks becomes increasingly skewed with shear, developing a low tail stretching past unity (Fig. \ref{fig:PdgProts}). It is not obvious why the shear has this effect, except to note that additional periodogram peaks appear with increasing significance in runs with shear, with a spreading of peak periods as the shear increases.

Period detection becomes significantly more inaccurate as the spot lifetime is reduced to two rotations. In that case only about 30\% of the solutions are within 20\% of the correct 1.0 period. A little over half of them have periods greater than 1.2, 40\% are greater than 2.0, and 15\% are greater than 3.0. This is a good illustration of why periodogram methods can fail on stars whose spots last only one or two rotations (the Sun is a member of that class). The addition of differential rotation doesn't make things much worse, nor does it improve them. Spots are coming and going too often to easily establish a solid periodicity. It is not that harmonics of the period confuse things; the dominant period itself is just often obscured. 

To drive this point home, we computed a few models where the spot lifetime is only one rotation. Those results are included in brown in the upper left panel of Fig. \ref{fig:PdgProts}. The four curves shown are for spot number 2 and 3 with no shear, as well as spot number 6 with zero or solar shear, but they all closely resemble each other and are equally poor. In this instance the method finds the right period hardly any of the time, tending to find periods anywhere between 1-3.5 times the true period. There is no way to decide what the real period should be based on these scattered results. At least in the context of our spot models there is no way that stars with short-lived spots like this will yield consistently accurate rotation periods from periodograms. It is worth pointing out that many spots on the Sun last less than one rotation period. It is thus no wonder that the Sun itself only occasionally yields its rotation period from an integrated light curve, and it is more often faculae on the quiet Sun rather than spots on the active Sun that produce the diagnostic signal.

\begin{figure}[H]
\includegraphics[width=\linewidth]{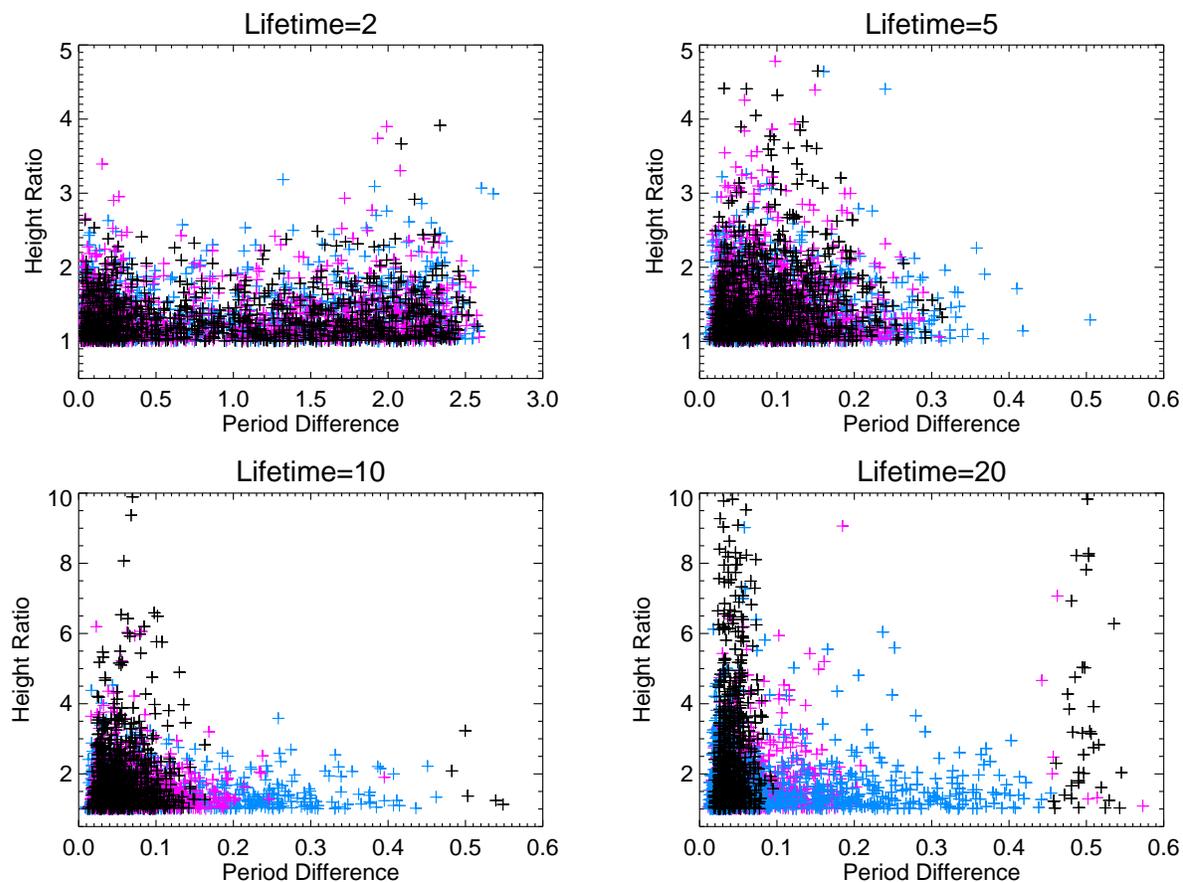}
\caption{ Periodogram Peak differences in time and height. The abscissa is the difference in time between the highest 2 periodogram peaks for each 6-spot run, and the ordinate is the ratio between the heights those two peaks. Black points are for zero shear, magenta points are for solar shear, and blue points are for twice solar shear.}
\label{fig:PdgHtDif}
\end{figure}

The other question that can be easily investigated is whether differential rotation leaves a signature in the periodograms that can help us to measure it. This was the idea employed by \citet{RRB13}, who interpreted the presence of two reasonably equal and high periodogram peaks near each other in period as a signature of shear. Given that the models can generate a light curve from the very same conditions with differential rotation turned on or off, we can test this interpretation. Fig. \ref{fig:PdgHtDif} shows the absolute value of the period difference between the top two peaks in a given case on the abscissa (the distributions are essentially symmetric about positive and negative differences). The ratio in height between them (top over second) is shown on the ordinate, for canonical models with zero, solar, and twice solar shear. If the conjecture above is right, we should find that points with low ratios (nearly equal heights) should cluster at small period differences as differential rotation is turned on, and models with no differential rotation should have greater height ratios (one main peak). The lifetime 20 case does show some of these characteristics. The zero shear case has small period differences for most trials, and a small set of trials at the first harmonic (0.5 period difference). Most of these trials have height ratios of 2 or more, meaning that the main peak is prominent. Turning on solar shear reduces the height differences, and produces greater period differences (mostly within 0.2 periods). Doubling that shear enhances both these effects, leaving many trials with low height ratios and larger period differences as anticipated by the original conjecture.

Unfortunately, this nice result already breaks down at lifetime 10, except for twice solar shear. The other cases cluster together below height ratio of 4 and period difference of 0.15, and there isn't much difference between the distributions with zero shear and solar shear. The effects are completely erased by lifetime 5, and lifetime 2 has period differences that range from small to 0.5 (without clustering at 0.5) and height ratios mostly less than 2. There are also often additional significant periodogram peaks within these limits. This means that stars with something resembling solar spot lifetimes (in rotation units) will completely defeat this method of looking for differential rotation. That is consistent with the results reported by \citet{Aig15}, and puts all reports of measurements of differential rotation from light curves in question unless there is good evidence that the spots are quite long-lived compared with the physical rotation period. Everything we have learned supports the idea that differential rotation can only be distinguished from spot evolution in such long-lived cases, and even then it also requires that spots be spread generously in latitude so as to sample different effects of the shear (as is true in the models).

\section{Discussion and Conclusions \label{sec:Conclusions}}

We have computed a large number of starspot models, varying the number of spots, their locations, their lifetimes, and the possible presence of latitudinal shear (differential rotation), as well as the inclination the observer sees. The models are intended to represent stars that are at least as active as the active Sun (and usually much more so). That is why our ``spots" are rather large, and better thought of as representing spot groups. That doesn't matter much, however, because the spatial resolution of the information from light curves is quite low. The light curves exhibit quite a range of appearances, similar to stars for which rotation periods have been determined in Kepler light curves, that could be interpreted with a number of different physical explanations. Yet the models in some sense are all similar: same-sized spots distributed randomly in space and time. The primary physical property that produces clear qualitative differences is the spot lifetime; spots that last only one rotation generate quite different light curves than those that last 50 rotations.

In this paper we have only begun to scratch the surface of possible starspot models. We use a constant value for the maximum spot size, a constant contrast, and usually a random distribution over the whole visible star as well as through time. We did not consider the effects of penumbrae or faculae. We do not present results with belts of spots confined to latitude ranges or concentrated near the pole or in active longitudes (although we did examine a few models for each of these cases). The results suggest that it may not be useful to add such details while we are ignorant of some basic stellar parameters in large datasets like those current precision photometry missions are producing. It would be very helpful to know at least the inclination and rotation period for each star. It is also very helpful to observe a large number of rotations, because the light curve may not exhibit statistically characteristic behavior over a few rotations. The problem is exacerbated when the light curves are only differential (when the absolute flux is not calibrated). The main aim of this paper is to help understand what information is and is not contained in such light curves, and how to extract it.

\subsection{Light Curve Degeneracies \label{sec:Degeneracies}}

We have presented a few illustrative examples of specific models, but more generally the histograms of important light curve metrics as a function of various physical parameters of the models. They show that it is very difficult to be certain of uniquely measuring any of the physical properties of starspot manifestations that are of interest. The behavior of the light curves is much simpler than the behavior of the underlying starspot distributions, primarily because the light curves sample the entire visible hemisphere at any given time. Starspots vary their influence either because they are physically growing/shrinking or because their projected area is changing as they pass closer to/further away from the sub-observer point. Their distribution on the stellar surface can also change because of either starspot evolution or differential rotation. 

Particularly instructive is the behavior of the fundamental measurable quantity in these light curves: the depth and duration of dips in differential intensity. There are generally only one or two such dips per rotation of the star, and the phase and duration of the dips can change with each rotation. This point has been made more than once in the past without gaining enough traction. A relatively recent reminder occurs in \citet{Jeff05}, who studied what could be called ``freckled" stars. The information contained in the intensity dips is primarily that of hemispheric asymmetries in the brightness of the part of the star that is sampled by the observer. The whole star is visible when viewed equatorially over one rotation, but the visible fraction drops toward half the star as the inclination goes to zero. On the next rotation spots may have changed their size or their position in stellar coordinates, and some may have disappeared while new ones have appeared. Depending on the phases of all the spots, one can often divide the star into a brighter and dimmer hemisphere. In that case the light curve will exhibit a single dip per rotation; this is the dominant situation in the large set of models we have computed. 

We also checked on the phasing of the dips by setting zero phase at the first light curve minimum then using the rotation period to define phase, particularly looking for consistency or regular drifting within extended single-dip segments. Unfortunately this did not turn out to be a promising avenue to pursue, as all sorts of behaviors occurred. It would be fair to say, however, that a consistent phase hardly ever lasted very long, except in the case of few very long-lived spots. We did an initial examination of how controlling the spot distribution in longitude shows up in the persistence of the dip phase. Excluding spots from a 90-degree slice of the star produces almost entirely single-dipped light curves, although the phase of the dips wanders. Adding solar shear returned the SDR to nearly the original distribution. Placing two permanent spots at zero longitude with the same size as the other spots in our canonical case (with lifetime 5) was not enough to force the dip phases to collect near zero, but making them four times bigger did cause about half the dips to occur near zero phase. Better constraints on detection of true ``active longitudes" will be pursued in future work.

There are also spot configurations that cannot be divided into a single pair of darker and brighter hemispheres; the light curve reverses and gets brighter then dips and brightens again during one rotation. This typically happens only once (a double dip) but can occasionally happen another time, usually with a small amplitude. The relative amplitudes of the double dips can be similar or quite different. It is sometimes the case that a small dip appears on one side of a larger dip and migrates toward the other side. One interpretation of this is the effect of differential rotation, as indeed it can be. Unfortunately we find that it can more easily arise because new spots are forming in the right place(s) (or old spots are disappearing). The differential rotation interpretation is only secure if spots don't evolve. We also show that the amplitudes of the light curve dips are not directly reflective of spot coverage. The dips will be smaller if the spot distribution is more symmetric and larger if more asymmetric; paradoxically it can be easier to get a larger asymmetry and differential amplitude with a smaller number of (sufficiently sized) spots. On the positive side, the variability metric does seem to respond in a usable way to the spot number, and this is even more true if differential rotation is present. It may be that we can infer spot numbers to within a factor of a few; the case presented here is the simplest so this is simply indicative at the moment.

Changes in the amplitude of the light curve variability do not necessarily indicate changes in coverage or activity level, they can also be reflective of simple (and relatively subtle) rearrangements of the spot distribution. Sometimes the dip durations display interesting and recognizable patterns, usually for the double dips. An example of this is seen in the middle panel of Fig. \ref{fig:dbsnwid}. The oval pattern between rotations 18-26 is indicative of alternation between smaller/larger dip durations, with the position of the smaller dip drifting from one side of the larger dip to the other. Another pattern seen is a flat distribution of dip durations near a half or whole period. The presence of such patterns in observed light curves indicates that the dips are real (not noise), but unfortunately they do not distinguish between the possible physical causes of that light curve behavior. There might be algorithms or applications of machine learning that recognize these patterns and help us understand the spot distributions. In many other cases, especially for short spot lifetimes, the dip durations are significantly more scattered and show little organized pattern.

We have presented the statistics of the relative occurrence of single- or double-dip light curve segments in the models using the SDR metric. We show that positive values of SDR (predominantly single-dipped segments) are obtained for most values and combinations of parameters. The best ways to push the SDR negative are to have short spot lifetimes and/or to view stars at high inclination. The distributions of model SDRs are more positive than those observed in real stars by Kepler \citep{Bas18a}. The peak of the distribution of SDRs for the main sequence stars in the \citet{McQ14} sample is at about -0.6 with a FWHM of 1.0, which means that nearly a third of the observed SDRs are less than -0.5 (below any model SDR). This is probably due to the fact that the observations contain noise (which produces non-physical dips in the light curve), while the models are noiseless. We make a preliminary guess that observed SDRs less than -0.25 or so are increasingly caused by noise, and that many of the observed SDRs are displaced by some amount to the negative, presumably more so for faint stars with low intrinsic ranges. To really demonstrate this one will have to run a set of models with a proper noise model added, which we leave for future work.

\subsection{Measurement of Spot Lifetimes \label{sec:SpotLifetimes}}

\begin{figure}[H]
\includegraphics[width=\linewidth, height=18cm]{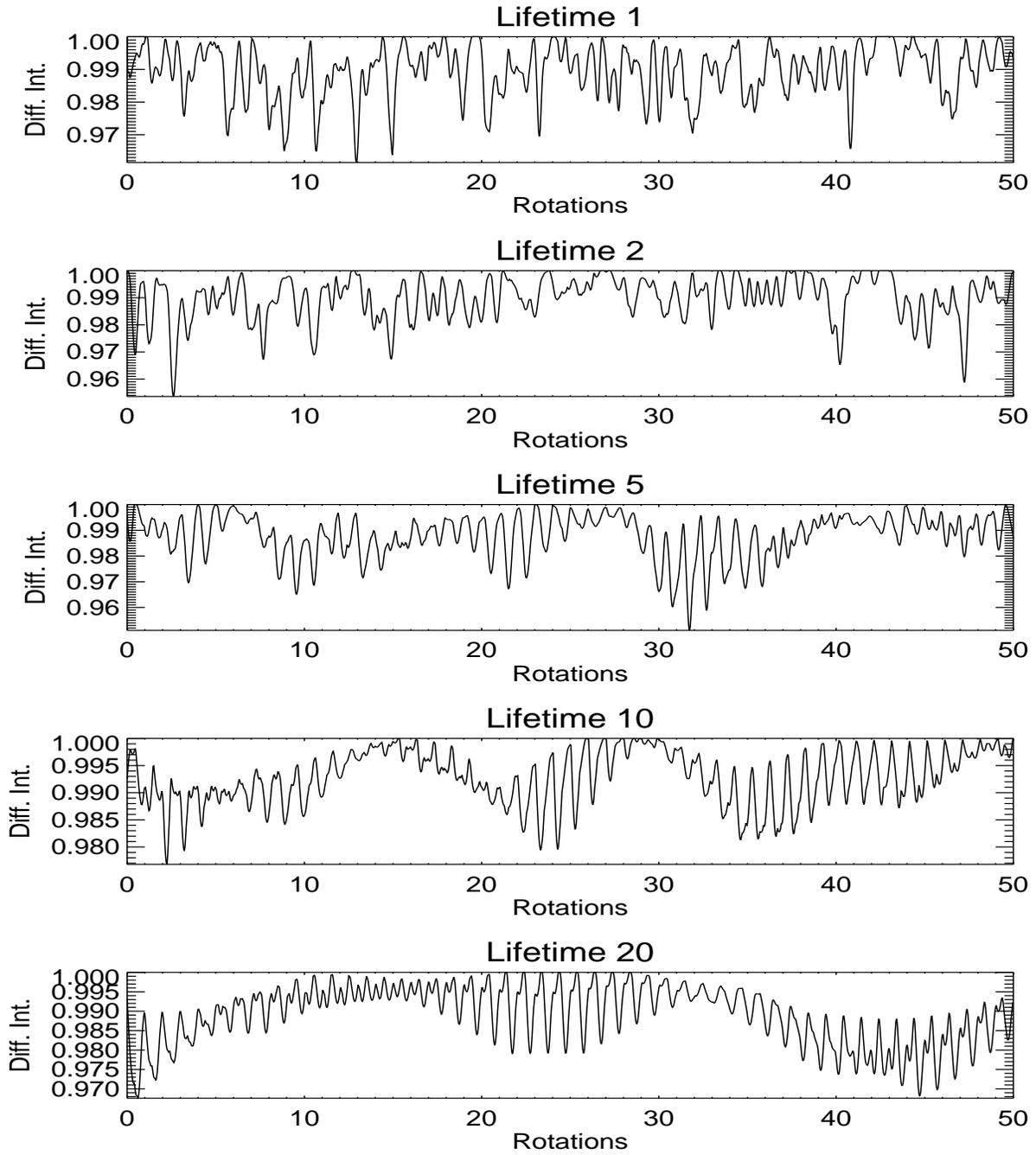}
\caption{ A random sample of light curves with a set of different spot lifetimes. The other parameters are all from our canonical set. }
\label{fig:LifetimeLCs}
\end{figure}

There is some hope that spot lifetimes can be determined to some accuracy, because our modeling shows that spot lifetime does produce distinguishable changes in the light curves. Some of these effects can be seen in Fig. \ref{fig:LifetimeLCs}. It is qualitatively clear that the longest lifetime case looks quite different from the shortest lifetime case. For lifetimes of 2 rotations or less, the light curves look fairly random, exhibiting few discernible repeating patterns. As described in Section \ref{sec:Periodograms}, that makes even the rotation period hard to measure well, but it does indicate a short spot lifetime. Increasing to lifetime 5 we begin to see ``short activity cycles" or what we have called ``variability valleys". Those light curves have more coherent segments of single or double dips, and the former have larger amplitude. The ``cycles" have lengths that are 2-3 times the spot lifetime, but are actually just stochastic variations. This same basic behavior becomes clearer at lifetime 10. At lifetime 20 the cycles are a smaller multiple of the spot lifetime, but this is partially because of the relatively small number of spots in play at any given time. These models use fixed spot lifetimes; presumably if a star has a mix of lifetimes the effects will become somewhat harder to detect (and the ``spot lifetime" is less well defined in that case).

We already began the discussion of how much to trust methods of finding spot lifetimes in Section 3.2 of \citet{Bas18a}. That discussion was about the analysis in \citet{Gil17}, who interpret the degradation of the autocorrelation function (ACF) at greater shifts to be an indicator of the spot distribution changing due to the appearance and disappearance of spots, and thus a measure of spot lifetime.  They interpret the changing variability as due to spot size changes rather than changes in spot positions, and regard the decay of the pattern repetition as diagnostic of starspot decay times. They are (appropriately) more cautious about interpreting the phase of the larger dips as active longitudes. \citet{Bas18a} agreed that spot evolution is one possibility for the degradation but pointed out that differential rotation could also cause the ACF to degrade due to changes in spot positions. Our results in this paper lead us to agree that spot evolution is the primary cause of ACF degradation except when spots have long lifetimes, where differential rotation has a better chance of producing the same effect (see Section \ref{sec:DiffRot}).

The results in Fig. \ref{fig:SDRdiffincl} can be viewed as generally supportive of the idea that both spot lifetime and differential rotation remain possible culprits in the changing light curve appearance. The main variable that shifts the SDR is stellar inclination, but that is obviously fixed for a given star. The SDR spreads to both larger and smaller values as the spot lifetime increases, and differential rotation inhibits that spread, but the distributions remain centered about similar values of SDR in all cases for a given inclination. It is reasonable to suppose that spot lifetimes are connected to physical variables like convection velocity. These will be similar for stars with similar basic stellar parameters, which might lead to similar spot lifetimes in days. In that case a spot that lives 20 days would live for 4 rotations on a star with a rotation period of 5 days, but less than one rotation for a similar star with a solar rotation period. Most of the stars in the Kepler sample are likely to have ``short" spot lifetimes measured in rotation periods, because most of them are closer in age to the Sun (or older) than to young rapid rotators. 

One important cause of the degradation of the ACF utilized by \citet{Gil17} is the switching between single and double dip modes; light curve segments certainly do not correlate as well with each other across such switches. We can test this reasoning by looking at how the ACF behaves in our models. We examine cases with the canonical spot number and inclination and no shear, but various spot lifetimes. As a simple measure of the ACF degradation we take the ratio of the height of the third ACF peak to the first one; the smaller than unity this ratio is the faster the autocorrelation is diminishing. 

\begin{figure}[H]
\includegraphics[width=\linewidth]{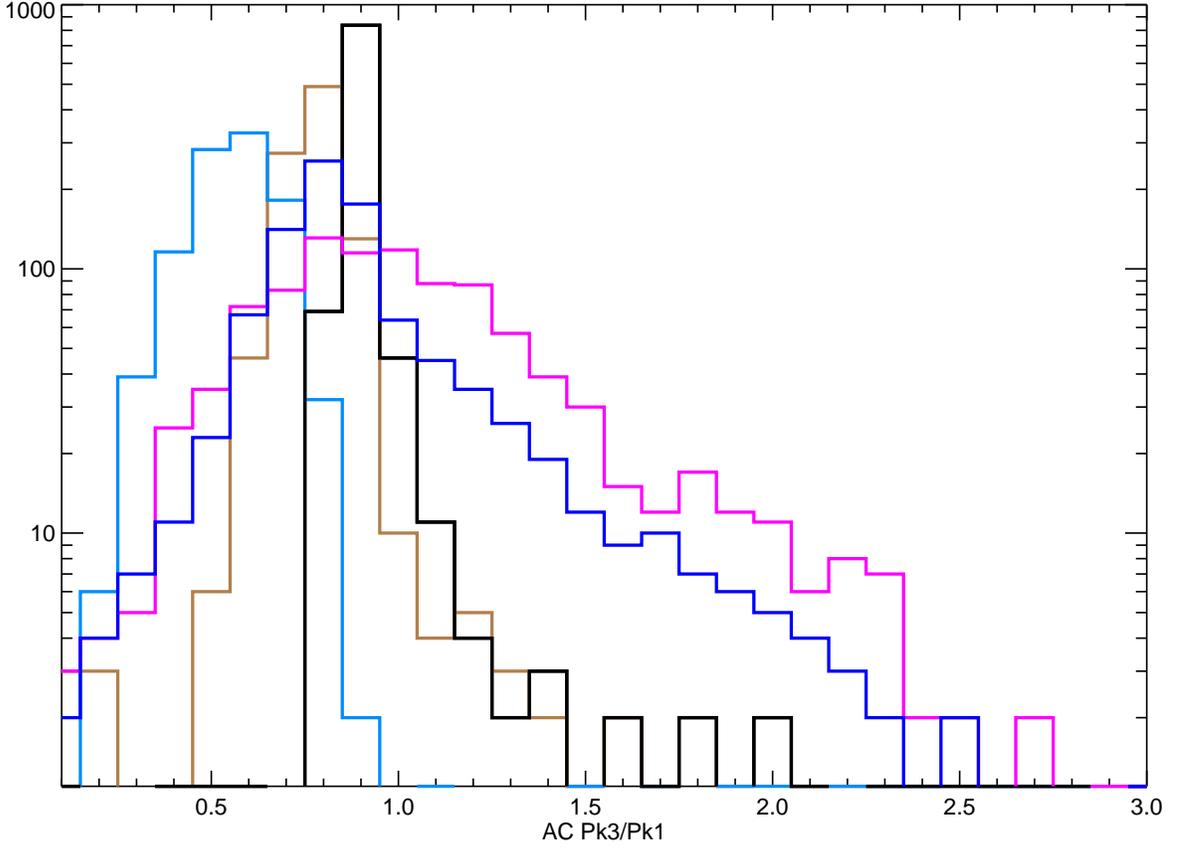}
\caption{ Histograms of autocorrelation peak ratios. The ratio of the height of the third AC peak to the first (a measure of how fast it is degrading) for various cases. The model cases (1000 trials each) have various lifetimes: black is 20, brown is 10, light blue is 5 and purple is 2. The main sequence stars in the MAM14 sample (15000; renormalized) are shown in dark blue. These results suggest that the spot lifetimes for most of the observed stars may be less than 2 rotation periods. }
\label{fig:ACPkRat}
\end{figure}

Fig. \ref{fig:ACPkRat} shows the results for lifetimes of 20, 10, 5, and 2 rotations. At 20 the ACF degrades hardly at all, the peak ratios are mostly near unity. This means that the pattern is preserved for several rotations, as might be expected. It is a little surprising that in a few cases (note that the vertical scale in Fig. \ref{fig:ACPkRat} is logarithmic) the peak ratio is significantly above unity. These turn out to be cases where the ACF has a very small peak followed by a large peak -- a classic signature of double dips, where the second peak is the one corresponding to the real rotation period. The third peak (second small peak) utilized in our ratio is somewhat larger than the first in these cases. For such cases our ratio is not ideal; we are missing the main peaks. At lifetime 10 there is some ACF degradation: the peak ratio distribution is highest at 0.8 and extends down to 0.5. At lifetime 5 there is more degradation, with the maximum at 0.6 and some cases as low as 0.2. An increasingly small number of cases remain near unity as the lifetimes drops from 10 to and there are no cases above unity. 

At lifetime 2 things get far more interesting and less predictable. In this case the majority of peak ratios are actually greater than unity, extending to more than twice that, meaning that the ACF is substantially higher for greater shifts away from zero in many cases. It is not clear why the correlation can grow several rotations out, although the general level of correlation is lower than for longer lifetimes. The light curves look quite random, with large and small dips well interspersed (Fig. \ref{fig:LifetimeLCs}). The spacing of the ACF peaks also becomes more erratic, and the peak shapes become increasingly irregular. Some peak ratios remain below unity, but clearly the degradation of the ACF for one given light curve (at least using our simple metric) is no longer predictive of the spot lifetime. On the other hand, the aggregate behavior is quite distinctive. Fig. \ref{fig:ACPkRat} also contains the same ACF degradation diagnostic applied to 15,000 main sequence stars in the MAM14 sample from Kepler. That result clearly resembles the lifetime 2 case, which implies that the observed sample consists largely of stars with spots whose lifetime is only one or two rotations, as is reasonable. This appears a promising line of analysis we will pursue in the future. It will make sense to define a more sophisticated metric for ACF degradation since the AC peaks are not always a clean descending set, especially for short spot lifetimes.

Another potential metric for spot lifetime, given the behavior of the light curves illustrated in Fig. \ref{fig:LifetimeLCs}, is the duration of the longest single-dip segment (assuming that one has a light curve with a lot of rotations). For our models with 100 rotations, the peak and FWHM of the distributions of this metric are: for lifetime 2 the peak is at 2 rotations and the FWHM is also 2; at lifetime 5 they are 7 and 8 respectively; for lifetime 10 they are 13 and 10 respectively; for lifetime 20 the peak is at 15 and the FWHM 16 (all in rotations). This shows that the peak of the distribution of the longest single segment increases with spot lifetime, but it is also true that the distributions increasingly overlap. It is therefore hard to assign a particular lifetime to a particular light curve, but if the longest single segment is more than 15 rotations it is fairly safe to say that the spot lifetime is greater than 5, and if it is less than 5 rotations the lifetime is likely less than 5. This is another promising form of inquiry into spot lifetimes that bears further development.

\subsection{Detection of Differential Rotation \label{sec:DiffRot}}

Although we tested the effects of differential rotation on a number of different light curve metrics, the effects are not very significant on the majority of metrics tested including SDR, coverage, range, and periodogram peak separations. Range and coverage are particularly insensitive; that is by design in the case of coverage. The only cases in which differential rotation changes the metrics significantly compared to no-shear cases are those in which the lifetime is quite long (at least 20 rotation periods). The general effect of differential rotation is to reduce the spread of the distribution of SDR. We originally thought that there would be a tendency for more double dips when shear is added, but that turns out not to be the case except for the longest-lived spots. There are not significant differences between differential rotation and no-shear cases when changing spot number; inclination has the same qualitative effects with and without shear. We showed in Section \ref{sec:Periodograms} that a periodogram analysis cannot diagnose differential rotation except for very long-lived spots. In that case the pattern changes in a systematic way that appears in the difference in time between the highest two periodogram peaks, as has been employed by some authors. 

\begin{figure}[H]
\includegraphics[width=\linewidth]{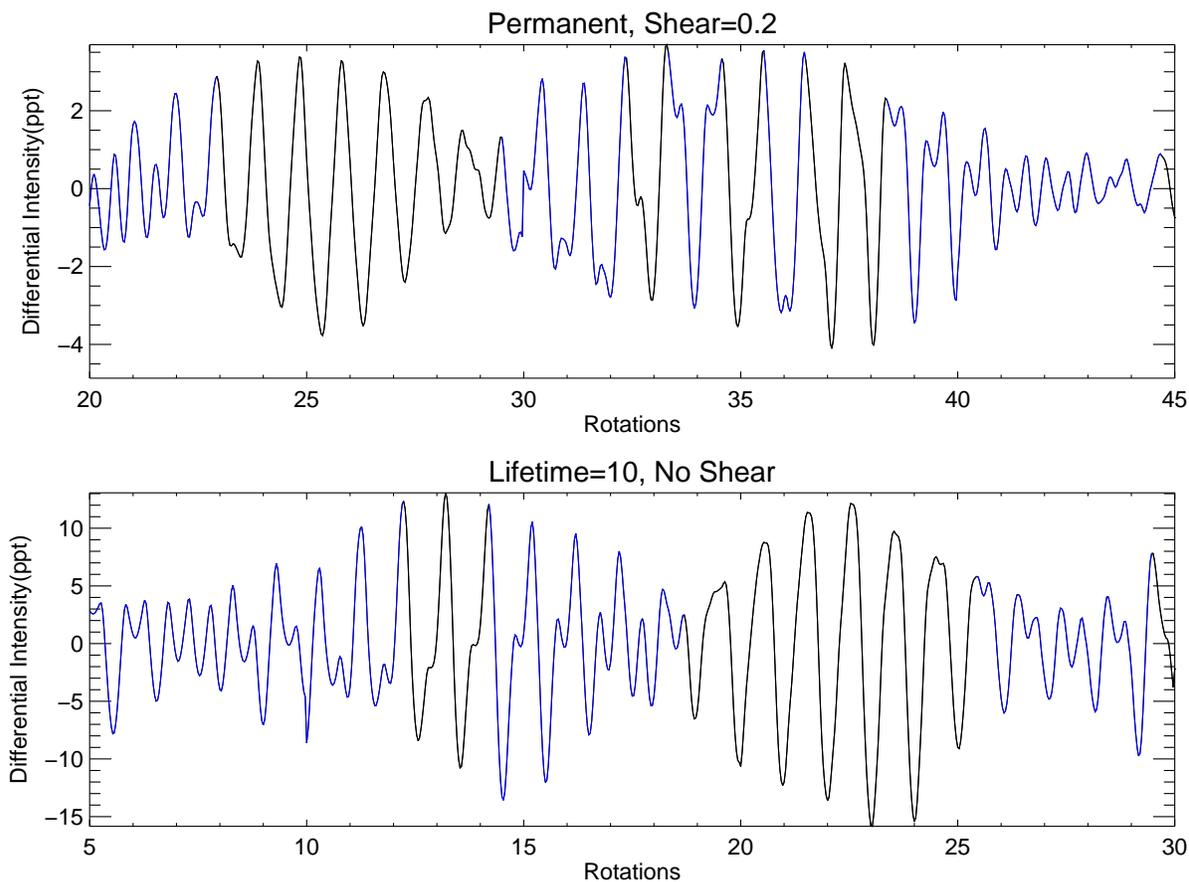}
\caption{ Sections of two light curves illustrating the difficulty in distinguishing between spot evolution and differential rotation. The upper curve has solar shear but no spot evolution, while the lower curve has spot evolution but no shear. Blue segments are double-dipped and black segments are single-dipped. These examples were selected to look particularly similar, but they demonstrate the difficulty in separating these two important physical behaviors {\it a priori}.}
\label{fig:EvDifRot}
\end{figure}

The real problem is that the effects of differential rotation only really become uniquely discernible if one has knowledge of other parameters, particularly spot lifetime and stellar inclination. The distributions of the light curve metrics overlap each other too much to allow the parameters to be uniquely determined by any given light curve. There are just too many ways to produce a light curve that is similar in these metrics. For illustration we show in Fig. \ref{fig:EvDifRot} portions of two light curves for 6 spots and $60^\circ $ inclination that closely resemble each other qualitatively. Both light curves have been ``Keplerized", meaning that secular drifts in absolute intensity are removed on a quarterly basis (a ``quarter" was set to 10 rotations) and the quarterly medians normalized to zero. The upper panel shows a case where the spots are permanent and there is solar shear. In the bottom panel the spots live for 10 rotations but there is no shear. It will be very difficult to devise metrics that can accurately distinguish between the two! The differential rotation case is essentially for very long-lived spots, however most stars do not have such long spot lifetimes. Thus patterns like those in Fig. \ref{fig:EvDifRot} are more likely to be due to spot evolution except perhaps on very rapid rotators. As mentioned above, there might be subtle changes in the pattern of dips that could be discovered if a large set of models is used as a training set in a machine learning approach. Then one might be able in a statistical sense to learn something about differential rotation. Unfortunately it could easily turn out that the inverse problem is simply too ill-posed.

It should be clear from Section \ref{sec:DiffRot} that differential rotation in most cases doesn't shuffle spots around in a way that can be convincingly and uniquely detected in a light curve by any of the strategies that have been employed so far. That means that all the qualitative behavior that is encompassed by those metrics (SDR, Range, Variability, variability ``cycles") is not influenced in a unique way by differential rotation. We discussed in Section \ref{sec:Periodograms} the use of multiple periodogram peaks. Papers using this include \citet{RRB13},  \citet{Rein15}, and \citet{Sant17} (the latter paper also mentions several others). The periodogram method only works well when the spot lifetime is greater than 10-15 rotation periods. That may be the case for some of the rapid rotators in their samples, and more likely is the case for very young stars (which the K2 mission has shown can display very persistent light curve patterns), but isn't the case for the large samples in the above papers. Such an active rapid rotator is GJ 1243, for which differential rotation was derived using an MCMC spot model by \citet{Dav15}. While this star has a better chance of showing patterns primarily due to differential rotation, it is still subject to the criticism above about the ability of few spot models to fit the light curve without strong correspondence to the actual physical situation. The same can be said of the approach of \citet{Bal16} who use a method akin to wavelet analysis, but this does not help with the underlying conceptual issue. 

Finally, \citet{Lanz14} and \citet{Das16} use an autocorrelation method combined with MCMC 2-spot models. We checked the analysis in \ref{sec:SpotLifetimes} to see whether differential rotation affected the ACF degradation that was discussed there. Not surprisingly, given that it doesn't affect the SDR, it also does not affect the peak ratio diagnostic we used. All these approaches are flawed for the reason that the changes in the amplitude, positions, and repeatability of the light curve dips are not necessarily, mostly, or even usually changing due to differential rotation.

Unfortunately, spot evolution can induce very similar behaviors in light curves, so unless there is a way to separately fix the spot lifetime it will be difficult to assign a confident value to the shear. The situation is considerably worsened by the fact that if solar-type stars behave like the Sun, spots are concentrated into relatively narrow latitudinal belts at any given time, so the shear is not well sampled by the spots present (whereas it is by construction in our models). It is therefore our contention that none of the claims so far of measurement of differential rotation based solely on precision photometric light curves is trustworthy (including the 2013 paper on which GB was an author). This is a great pity, because those are the only papers that include large samples of stars with a good spread of stellar parameters. We are not willing to completely give up on the quest for such determinations yet, the datasets are rich and tempting, but deeper work will be required to extract confident measurements of differential rotation.

\subsection{Implications for other work \label{sec:Implications}}

In this paper we have expounded a more detailed understanding of differential light curves and the underlying spot distributions that produce them than has usually been implicit in work in this area. It has important implications for a significant number of previous papers analyzing precision light curves. We end by providing a few examples on two more topics where it makes sense to re-think the interpretations that have been drawn in the past, or at least to retain a healthy skepticism about their veracity. We presume for this discussion that the correct stellar rotation period has been pre-determined with sufficient accuracy (perhaps within a spread due to differential rotation); without that one is truly lost. We do not present an exhaustive list of papers on these topics, and we do not intend to impugn the work in the cited papers, nor in the papers mentioned in the sections above. Our purpose is rather to encourage future authors to take the considerations presented here into account. 

The most obvious and widespread misinterpretation is that of single and double light curve dips per rotation as one or two spots (or spot groups, or even just active longitudinal regions) on the star. A recent example of this is found in \citet{Sav16}, who provide two-spot maps of an M dwarf and also discuss the drift of an ``active longitude". That paper carries on a tradition of such analyses stretching back to the 1980s; an older example is \citet{Stras94}. Such solutions can be well-posed, and sometimes analytic or semi-analytic techniques can generate them. In modern days it is easy to solve this sort of problem with an MCMC approach. Recent papers using that approach include \citet{Ion16} and \citet{Name19}. In each case a good fit to the light curve can be achieved by using around 4 spots, but it is not at all clear how closely the solution corresponds to how the star actually looks. \citet{Name19} interpret the evolution of single dips as direct indicators of the lifetime of the spot that is causing the dip; the results here clearly argue for re-interpretation of those lifetimes as only signifying how long a particular asymmetry in the spot distribution lasts (for whatever reasons). 

In a paper which strongly empirically supports our main point, \citet{Name20} take advantage of data from a star (Kepler-17) where a transiting planet actually passes over some spotted regions, allowing their direct and individual measurement. They show that there often must also be a lot of other spots on the star that generate the observed light curve, and the combination of seen and unseen spots doesn't correspond well to the 4-spot MCMC solution that generates a good fit to the light curve. One must therefore be very careful about what is claimed from such solutions, beyond the asymmetry of the spot distribution and a lower limit on the amount of spot coverage. The changes in rotation phase of the dip(s) are really only a measure of where the asymmetries are strongest over time, localized to perhaps a quarter of the stellar surface. The depth, duration, and timing of the dips have an obscured relation to the number of actual spots, and depend sensitively on accidents of positioning and current observed sizes. \citet{Lanz19} also analysed this star and implicitly reached similar conclusions, although they concentrate on what the interpretation would be if one believed the ``active longitude" metaphor for the light curve.

Another type of study where there often is misinterpretation of the single/double behavior and its accompanying changes in variability is the search for activity cycles in light curves. We have shown that the variability can be strongly affected by changes in the spot distribution that do not reflect changes in the coverage. Generally single-dipped segments have about twice the variability of double-dipped segments \citep{Bas18a}; this is because they are reflective of larger-scale asymmetries in the spot distribution. Changes in coverage can also occur due to random fluctuations in the appearance of spots, and if there are not that many spot groups present on average these fluctuations can look like significant changes in activity. Depending on the spot lifetime, they can look like short activity cycles. It is a result of the relatively short duration of the Kepler mission (4 years) that activity cycles that have been claimed from mission data are all relatively short, but that opens them to doubt in light of our models. 

One example of a paper that uses a method subject to this criticism is \citet{Vida14}. Those authors pass a Gaussian filter with a width that is tuned to be sensitive to changes in the Fourier spectrum of the selected region; that is ideally suited to focus in on changes between single- and double-dipped regions. They detect ``cycles" of a few months, which is also characteristic of light curves that switch modes a few times during a hundred rotations (if the rotation periods are within a factor of two of a month). Other papers that study essentially the same sort of changes in variability that could arise simply from the random appearance in time and location of spots with moderate lifetimes include \citet{Ark15}, \citet{Rein17}, and \citet{Niel18}. It is important to note that just because these longer term variations can arise through causes that would not generally be considered activity cycles, we also cannot be certain that they are due to random variations. As with many of the points in this paper, the takeaway is that more skepticism is warranted and further analysis is required, not that the claimed results are definitely incorrect. 

This paper only addresses what can be learned about starspots from broad-band light curves. Activity cycles can in principle be seen in light curves but such an interpretation has to be made more carefully than to date. There is certainly clear evidence for (typically longer) activity cycles in other diagnostics (such as Ca II or X-rays). The addition of color information improves matters, and spectroscopy even more so. Furthermore there are several other techniques, including Doppler imaging, Zeeman Doppler imaging, and transiting of giant exoplanets over spots, that are not subject to the set of considerations relevant to this paper. Of course, there is much less available data in these more informative modes. The analysis of how the issues we have raised here actually affect these other techniques will be complicated, and contingent on exactly what sort of observations are being evaluated. No method is entirely free of the problem of effective spatial resolution.

In summary, this paper strongly reinforces the long known but insufficiently applied point that the extraction of starspot information from precision broad-band light curves is an attempt to learn physical information from a very degenerate and ill-posed inverse problem. It does so by illustration from forward modeling of slightly more realistic cases than usually considered. The simplifications that were (perhaps) adequate when the quantity and quality of stellar light curves was far less than what is available now will simply no longer do. Researchers should be careful with language or models that suggest that one or two intensity dips per rotation in a light curve can confidently be represented by small numbers of ``spots" and/or ``active longitudes" unless they have independent evidence for it. We advocate that the language be changed to the single/double dip language used in this paper. If authors insist on the old simplifications, it should only be for inferences that do not depend on reasonably accurate spot distributions and are insensitive to the variety of configurations that can produce similar light curve changes. Thought should be given to whether such results can be subjected to the same criticisms as above.

It certainly seems that light curves such as are in the Kepler archive are rich with information, so we do not advocate throwing up our hands and saying that the interesting information is irretrievable (yet). Rather, it is an interesting challenge to develop new techniques that are able to more fully define what can and cannot be determined, and thus put the study of stellar magnetic activity with these techniques on firmer footing. There will continue to be a great many light curves collected for hundreds of thousands of stars for at least the next decade (TESS is doing so right now). The primary Kepler dataset will probably remain unique in containing the longest continuous monitoring of a great many stars. Together with models of sufficient sophistication and newly developed techniques of analysis, it provides the best means of cracking this difficult problem and advancing our knowledge of the astrophysics of stellar magnetic fields.

\acknowledgments

GB is grateful for illuminating discussions with A. Reiners, T. Reinhold, and A. Lanza, and appreciates support for this research from UC Berkeley.

{\it Facilities:} \facility{Kepler}, \facility{MAST}.

\appendix

\end{document}